\PassOptionsToPackage{unicode}{hyperref}
\PassOptionsToPackage{hyphens}{url}
\PassOptionsToPackage{dvipsnames,svgnames,x11names}{xcolor}
\documentclass[
]{article}

\usepackage{amsmath,amssymb}
\usepackage{lmodern}
\usepackage{iftex}
\ifPDFTeX
  \usepackage[T1]{fontenc}
  \usepackage[utf8]{inputenc}
  \usepackage{textcomp} 
\else 
  \usepackage{unicode-math}
  \defaultfontfeatures{Scale=MatchLowercase}
  \defaultfontfeatures[\rmfamily]{Ligatures=TeX,Scale=1}
\fi
\IfFileExists{upquote.sty}{\usepackage{upquote}}{}
\IfFileExists{microtype.sty}{
  \usepackage[]{microtype}
  \UseMicrotypeSet[protrusion]{basicmath} 
}{}
\makeatletter
\@ifundefined{KOMAClassName}{
  \IfFileExists{parskip.sty}{%
    \usepackage{parskip}
  }{
    \setlength{\parindent}{0pt}
    \setlength{\parskip}{6pt plus 2pt minus 1pt}}
}{
  \KOMAoptions{parskip=half}}
\makeatother
\usepackage{xcolor}
\ifx\paragraph\undefined\else
  \let\oldparagraph\paragraph
  \renewcommand{\paragraph}[1]{\oldparagraph{#1}\mbox{}}
\fi
\ifx\subparagraph\undefined\else
  \let\oldsubparagraph\subparagraph
  \renewcommand{\subparagraph}[1]{\oldsubparagraph{#1}\mbox{}}
\fi

\usepackage{color}
\usepackage{fancyvrb}

\DefineVerbatimEnvironment{Highlighting}{Verbatim}{commandchars=\\\{\}}
\usepackage{framed}
\definecolor{shadecolor}{RGB}{241,243,245}
\newenvironment{Shaded}{\begin{snugshade}}{\end{snugshade}}

\newcommand{\AttributeTok}[1]{\textcolor[rgb]{0.40,0.45,0.13}{#1}}

\newcommand{\CommentTok}[1]{\textcolor[rgb]{0.37,0.37,0.37}{#1}}

\newcommand{\ConstantTok}[1]{\textcolor[rgb]{0.56,0.35,0.01}{#1}}

\newcommand{\DecValTok}[1]{\textcolor[rgb]{0.68,0.00,0.00}{#1}}
\newcommand{\DocumentationTok}[1]{\textcolor[rgb]{0.37,0.37,0.37}{\textit{#1}}}

\newcommand{\FloatTok}[1]{\textcolor[rgb]{0.68,0.00,0.00}{#1}}
\newcommand{\FunctionTok}[1]{\textcolor[rgb]{0.28,0.35,0.67}{#1}}

\newcommand{\NormalTok}[1]{\textcolor[rgb]{0.00,0.23,0.31}{#1}}

\newcommand{\OtherTok}[1]{\textcolor[rgb]{0.00,0.23,0.31}{#1}}

\newcommand{\SpecialCharTok}[1]{\textcolor[rgb]{0.37,0.37,0.37}{#1}}

\newcommand{\StringTok}[1]{\textcolor[rgb]{0.13,0.47,0.30}{#1}}

\usepackage{longtable,booktabs,array}
\usepackage{calc} 
\usepackage{etoolbox}
\makeatletter
\patchcmd\longtable{\par}{\if@noskipsec\mbox{}\fi\par}{}{}
\makeatother
\IfFileExists{footnotehyper.sty}{\usepackage{footnotehyper}}{\usepackage{footnote}}
\makesavenoteenv{longtable}
\usepackage{graphicx}
\makeatletter
\def\maxwidth{\ifdim\Gin@nat@width>\linewidth\linewidth\else\Gin@nat@width\fi}
\def\maxheight{\ifdim\Gin@nat@height>\textheight\textheight\else\Gin@nat@height\fi}
\makeatother
\setkeys{Gin}{width=\maxwidth,height=\maxheight,keepaspectratio}
\makeatletter
\def\fps@figure{htbp}
\makeatother
\newlength{\cslhangindent}
\newlength{\csllabelwidth}
\newlength{\cslentryspacingunit} 
\newenvironment{CSLReferences}[2] 
 {
  \setlength{\parindent}{0pt}
  \ifodd #1
  \let\oldpar\par
  \def\par{\hangindent=\cslhangindent\oldpar}
  \fi
  \setlength{\parskip}{#2\cslentryspacingunit}
 }%
 {}
\usepackage{calc}

\usepackage[noblocks]{authblk}

\makeatletter
\@ifpackageloaded{caption}{}{\usepackage{caption}}
\AtBeginDocument{%
\ifdefined\contentsname
  \renewcommand*\contentsname{Table of contents}
\else
  \newcommand\contentsname{Table of contents}
\fi
\ifdefined\listfigurename
  \renewcommand*\listfigurename{List of Figures}
\else
  \newcommand\listfigurename{List of Figures}
\fi
\ifdefined\listtablename
  \renewcommand*\listtablename{List of Tables}
\else
  \newcommand\listtablename{List of Tables}
\fi
\ifdefined\figurename
  \renewcommand*\figurename{Figure}
\else
  \newcommand\figurename{Figure}
\fi
\ifdefined\tablename
  \renewcommand*\tablename{Table}
\else
  \newcommand\tablename{Table}
\fi
}
\@ifpackageloaded{float}{}{\usepackage{float}}
\floatstyle{ruled}
\@ifundefined{c@chapter}{\newfloat{codelisting}{h}{lop}}{\newfloat{codelisting}{h}{lop}[chapter]}
\floatname{codelisting}{Listing}

\makeatother
\makeatletter
\@ifpackageloaded{caption}{}{\usepackage{caption}}
\@ifpackageloaded{subcaption}{}{\usepackage{subcaption}}
\makeatother
\makeatletter
\@ifpackageloaded{tcolorbox}{}{\usepackage[many]{tcolorbox}}
\makeatother
\makeatletter
\@ifundefined{shadecolor}{\definecolor{shadecolor}{rgb}{.97, .97, .97}}
\makeatother
\ifLuaTeX
  \usepackage{selnolig}  
\fi
\IfFileExists{bookmark.sty}{\usepackage{bookmark}}{\usepackage{hyperref}}
\IfFileExists{xurl.sty}{\usepackage{xurl}}{} 
\hypersetup{
  pdftitle={An introduction and tutorial to model-based clustering in education via Gaussian mixture modelling},
  pdfauthor={Luca Scrucca; Mohammed Saqr; Sonsoles López-Pernas; Keefe Murphy},
  colorlinks=true,
  linkcolor={blue},
  filecolor={Maroon},
  citecolor={Blue},
  urlcolor={Blue},
  pdfcreator={LaTeX via pandoc}}

\title{An introduction and tutorial to model-based clustering in
education via Gaussian mixture modelling}

\author[1,*]{Luca Scrucca}
\author[2]{Mohammed Saqr}
\author[2]{Sonsoles López-Pernas}
\author[3]{Keefe Murphy}

\affil[1]{Università degli Studi di Perugia}
\affil[2]{School of Computing, University of Eastern Finland}
\affil[3]{Department of Mathematics and Statistics, Hamilton Institute,
Maynooth University}
\affil[*]{Corresponding author: Luca Scrucca, luca.scrucca@unipg.it}

\date{}
\begin{document}
\maketitle
\begin{abstract}
Heterogeneity has been a hot topic in recent educational literature.
Several calls have been voiced to adopt methods that capture different
patterns or subgroups within students' behavior or functioning. Assuming
that there is ``an average'' pattern that represents the entirety of
student populations requires the measured construct to have the same
causal mechanism, same development pattern, and affect students in
exactly the same way. Using a person-centered method (Finite Gaussian
mixture model or latent profile analysis), the present tutorial shows
how to uncover the heterogeneity within engagement data by identifying
three latent or unobserved clusters. This chapter offers an introduction
to the model-based clustering that includes the principles of the
methods, a guide to choice of number of clusters, evaluation of
clustering results and a detailed guide with code and a real-life
dataset. The discussion elaborates on the interpretation of the results,
the advantages of model-based clustering as well as how it compares with
other methods.
\end{abstract}
\ifdefined\Shaded\renewenvironment{Shaded}{\begin{tcolorbox}[interior hidden, enhanced, borderline west={3pt}{0pt}{shadecolor}, breakable, sharp corners, boxrule=0pt, frame hidden]}{\end{tcolorbox}}\fi

\newcommand{\x}{\boldsymbol{x}}
\newcommand{\X}{\boldsymbol{X}}
\newcommand{\y}{\boldsymbol{y}}
\newcommand{\e}{\boldsymbol{e}}
\newcommand{\xbar}{\bar{x}}
\newcommand{\ybar}{\bar{y}} 
\newcommand{\I}{\boldsymbol{I}}
\newcommand{\V}{\boldsymbol{V}}
\newcommand{\0}{\boldsymbol{0}}
\newcommand{\1}{\boldsymbol{1}}
\newcommand{\Sb}{\boldsymbol{S}}
\newcommand{\betab}{\boldsymbol{\beta}}
\newcommand{\Normal}{\mathcal{N}}
\newcommand{\T}{{}^{\!\top}}
\newcommand{\diag}{\text{diag}}
\newcommand{\df}{\text{df}}
\newcommand{\C}{\mathcal{C}}
\newcommand{\M}{\mathcal{M}}

\renewcommand{\hat}[1]{\widehat{#1}}
\newcommand{\thetab}{\boldsymbol{\theta}}
\newcommand{\mub}{\boldsymbol{\mu}}
\newcommand{\Sigmab}{\boldsymbol{\Sigma}}
\newcommand{\Lambdab}{\boldsymbol{\Lambda}}
\newcommand{\U}{\boldsymbol{U}}
\newcommand{\Deltab}{\boldsymbol{\Delta}}
\newcommand{\Prior}{\mathcal{P}}
\newcommand{\coldim}{d}
\newcommand{\invWishart}{\mathcal{IW}}
\newcommand{\priorShrink}{\kappa_{\scriptscriptstyle \Prior}}
\newcommand{\priorMean}{\mu_{\scriptscriptstyle \Prior}}
\newcommand{\priorMeanb}{\mub_{\scriptscriptstyle \Prior}}
\newcommand{\priorDOF}{\nu_{\scriptscriptstyle \Prior}}
\newcommand{\priorScaleIW}{\Lambdab_{\scriptscriptstyle \Prior}}
\newcommand{\priorScaleIG}{\varsigma^2_{\scriptscriptstyle \Prior}}

\hypertarget{introduction}{%
\section{Introduction}\label{introduction}}

Statistical research is commonly performed with variable-centered
methods using a sample from the population to devise central tendency
measures or an ``average'' (i.e., mean or median). The average is
assumed to represent the population under study and therefore, could be
generalized to the population at large (Howard and Hoffman 2018;
Hickendorff et al. 2018). Put another way, the statistical findings of
variable-centered methods are thought to apply to all learners in the
same way. In doing so, variable-centered methods ignore the individual
differences that are universal across all domains of human function
(Saqr et al. 2023). Learners are not an exception, they vary in their
behavior, attitude and dispositions, and they rarely ---if at all---
conform to a common pattern or an average behavior (Hickendorff et al.
2018; Törmänen et al. 2022). An ``average'' is thus a poor
simplification of learners' heterogeneity; consequently, methods to
capture individual differences have started to gain popularity with the
increasing attention to patterns and differences among students.

Person-centered methods can be broadly grouped into two categories;
heuristic, distance-based clustering algorithms (e.g., agglomerative
hierarchical clustering, and partitional clustering algorithms like
\(k\)-means) on one hand and \emph{model-based} clustering (MBC)
approaches (e.g., Gaussian mixture models, latent class or profile
analysis) on the other. Though we focus here on the MBC paradigm, we
note that ---contrary to variable-centered methods--- all
person-centered methods are generally concerned with the modelling of
heterogeneity by capturing the latent (e.g., unobserved or hidden)
patterns within the data into subgroups of homogeneous ``clusters'' or
``profiles'' (Howard and Hoffman 2018; Hickendorff et al. 2018).
Modelling the unobserved patterns within the data could reveal the
qualitative differences between learners. For instance, where students
may have different patterns of approaching their learning, capturing
such patterns would make sense as each different approach may benefit
from a certain course design, set of instructional methods, or
scaffolding approach (Hickendorff et al. 2018). Similarly, dispositions
such as engagement, motivation, and achievement are multidimensional and
vary across students; capturing such differences requires a method that
could handle such nonlinear multidimensional dispositions and identify
the different patterns.

This chapter deals with one of the person-centered methods; namely,
latent profile analysis (from the perspective of model-based clustering
via finite Gaussian mixture models). This represents a
\emph{probabilistic} approach to statistical unsupervised learning that
aims at discovering clusters of observations in a data set (Fraley and
Raftery 2002). We also offer a walkthrough tutorial for the analysis of
a data set on school engagement, academic achievement, and
self-regulated learning using the popular \textbf{mclust} package
(Fraley, Raftery, and Scrucca 2023) for R (R Core Team 2023) which
implements the approach. Whereas mixture models and \textbf{mclust} have
received growing attention within social science, they have not garnered
widespread utilisation in the field of educational research and their
adoption in learning analytics research has been relatively scarce.

\hypertarget{literature-review}{%
\subsection{Literature review}\label{literature-review}}

While examples of mixture models being applied in educational research
settings are relatively scarce compared to other methods of clustering,
some notable examples exist that address patterns in students' online
learning, patterns in students' disposition, or collaborative roles.

Most studies in education research that applied mixture models used
latent profile analysis (LPA) to identify students' profiles from
self-reported data. For example, Yu et al. (2022) performed LPA on a
data set of 318 students' survey responses about emotional
self-efficacy, motivation, self-regulation, and academic performance,
and identified four profiles: ``low'', ``average'', ``above average with
a low ability to handle the emotions of others'' and ``high emotional
self-efficacy''. In the work by Cheng, Huang, and Hebert (2023), the
authors analyzed 615 vocational education students' achievement emotions
in online learning environments, and found three profiles: ``blends of
negative emotions'', ``nonemotional'', and ``pure positive emotion''.
Hoi (2023) employed LPA on self-report data on classroom engagement from
413 first-year university students in Vietnam and found three profiles:
``highly engaged'', ``moderately engaged'', and ``minimally engaged''.
Scheidt et al. (2021) collected survey responses from 2,339 engineering
undergraduates about 28 noncognitive and affective factors using a
survey instrument and using Gaussian mixture models found four very
distinct profiles of students.

The use of mixture models to analyze online trace log data ---which is
at the core of learning analytics data--- is not nearly as common. In
the study by Zhang et al. (2023), the authors applied LPA to variables
related to debugging derived from students' programming problems
submission traces. They found a profile with higher debugging accuracy
and coding speed, another profile with lower debugging performance in
runtime and logic errors, and a third profile with lower performance in
syntactic errors who tended to make big changes in every submission.
Studies covering online collaborative learning are even more scarce. A
rare example is the study by Saqr and López-Pernas (2022), in which the
authors used latent profile analysis to identify students' roles in
collaboration based on their centrality measures. The mixture models
identified three collaborative roles that represented a distinct pattern
of collaboration: leaders, who contribute the most to the discussion,
whose arguments are more likely to spread; mediators, who bridge others
and moderate the discussions; as well as isolates, who are socially idle
and do not contribute to the discussion.

A considerable number of studies that applied mixture models further
investigate the association between profile membership and academic
achievement. For example, in the aforementioned study by Yu et al.
(2022), students with high emotional self-efficacy had higher academic
performance than the other profiles. In the study by Zhang et al.
(2023), the authors found that higher debugging accuracy was related to
higher scores in all exams, whereas there were no differences between
the two other identified profiles. By the same token, researchers have
attempted to find reasons why a certain profile emerged, or what are the
variables that are more associated with one profile more than the other.
For example, Hoi (2023) found that peer support, provision of choice,
and task relevance are the factors more likely to predict classroom
engagement profile membership. Yu et al. (2022) found that
self-regulation and motivation played significant roles in determining
profile membership.

Clearly, there are plenty of opportunities for further exploration and
investigation in this area that could augment our knowledge of learning,
learners' behavior, and the variabilities of learning processes
(Hickendorff et al. 2018). This is especially true given the numerous
advantages of the MBC paradigm over more traditional, heuristic
clustering algorithms, which we imminently describe. Subsequently, in
the rest of this chapter we elaborate on the theoretical underpinnings
of the family of Gaussian parsimonious clustering models implemented in
the \textbf{mclust} R package and additionally explore some advanced
features of the package, which we employ in an analysis of a real
educational research application thereafter. Finally, we conclude with a
brief discussion.

\hypertarget{model-based-clustering}{%
\section{Model-based clustering}\label{model-based-clustering}}

As stated above, clustering methods, in a general sense, are used to
uncover group structure in heterogeneous populations and identify
patterns in a data set which may represent distinct subpopulations.
While there is no universally applicable definition of what constitutes
a cluster (Hennig 2015), it is commonly assumed that clusters should be
well separated from each other and cohesive in an ideal analysis
(Everitt et al. 2011). Conversely, objects within a cluster should be
more similar to each other in some sense, in such a way that an
observation has a defined relationship with observations in the same
cluster, but not with observations from other clusters.

Traditional clustering approaches, like the aforementioned \(k\)-means
algorithm, and agglomerative hierarchical clustering, use distance-based
heuristics to produce a ``hard'' partition of cases into groups, such
that each observation is associated with exactly one cluster only. As
such approaches are not underpinned by a statistical model, assessment
of the optimal number of clusters is often a fraught task, lacking the
guidance of principled statistical model selection criteria.

Conversely, the MBC paradigm assumes that data arise from a (usually
finite) mixture of probability distributions, whereby each observation
is assumed to be generated from a specific cluster, characterised by an
associated distribution in the mixture (McLachlan and Peel 2000).
Ideally, mixtures of distributions are supposed to provide a good model
for the heterogeneity in a data set; that is, once an observation has
been assigned to a cluster, it is assumed to be well-represented by the
associated distribution. As such, MBC methods are based on a formal
likelihood and seek to estimate parameters (e.g., means, variances, and
covariances, which may or may not differ across groups) which best
characterise the different distributions. Rather than yielding only a
``hard'' partition, each observation is assigned a probability of being
associated with each mixture component ---such that observations can
have non-negative association with more than one cluster--- from which a
hard partition can be constructed. These probabilities are treated as
weights when estimating the component parameters, which brings the
advantages of minimising the effect of observations lying near the
boundary of two natural clusters (e.g., a student with an ambiguous
learning profile) and being able to quantity the uncertainty in the
cluster assignment of a particular observation to provide a sense of
cases for which further investigation may be warranted. Compared to
other approaches, the other main advantages of this statistical
modelling framework are its ability to use statistical model selection
criteria and inferential procedures for evaluating and assessing the
results obtained.

Inference for finite mixture models is routinely achieved by means of
the expectation-maximisation (EM) algorithm (Dempster, Laird, and Rubin
1977), under which each observation's component membership is treated as
a ``missing'' latent variable which must be estimated. This formulation
assumes that the data are \emph{conditionally} independent and
identically distributed, where the conditioning is with respect to a
latent variable representation of the data in which the latent variable
indicates cluster membership. Given the relative familiarity of latent
class and latent profile terminology in the social sciences, we now
explicitly cast MBC methods in the framework of latent variable
modelling.

\hypertarget{latent-variable-models}{%
\subsection{Latent variable models}\label{latent-variable-models}}

Latent variable models are statistical models that aim to explain the
relationships between observed variables by introducing one or more
unobserved or latent variables. The idea behind latent variable models
is that some of the underlying constructs or concepts we are interested
in cannot be measured directly, but only through their effects on
observable variables. Latent variable modelling has a relatively long
history, dating back from the measure of general intelligence by factor
analysis (Spearman 1904), to the structural equation modelling approach
(Jöreskog 1970), from topic modelling, such as the latent Dirichlet
allocation (LDA) algorithm (Blei, Ng, and Jordan 2003), to hidden Markov
models for time series (Zucchini, MacDonald, and Langrock 2016). Latent
variable models are widely used in various fields, including psychology,
sociology, economics, and biology, to name a few. They are particularly
useful when dealing with complex phenomena that cannot be easily
measured or when trying to understand the underlying mechanisms that
drive the observed data.

When discussing latent variable modelling, it is useful to consider the
\emph{taxonomy} presented by Bartholomew, Knott, and Moustaki (2011).
This can be particularly helpful, as the same models are sometimes
referred to by different names in different scientific disciplines.
Bartholomew, Knott, and Moustaki (2011, Table 1.3) considered a
cross-classification of latent variable methods based on the type of
variable (manifest or latent) and its nature (metrical or categorical).
If both the manifest and latent variables are metrical, the model is
called a \textbf{factor analysis model}. If the manifest variables are
categorical and the latent variables are metrical, the model is called a
\textbf{latent trait model} or \textbf{item response theory model}. If
the manifest variables are metrical and the latent variables are
categorical, the model is called a \textbf{latent profile analysis
model}. If both the manifest and latent variables are categorical, the
model is called a \textbf{latent class model}.

In this scheme, finite Gaussian mixture models described in this chapter
assume that the observed variables are continuous and normally
distributed, while the latent variable, which represents the cluster
membership of each observation, is categorical. Therefore, Gaussian
mixtures belong to the family of latent profile analysis models. This
connection is made apparent by the \textbf{tidyLPA} R package (Rosenberg
et al. 2018), which leverages this equivalence to provide an interface
to the well-known \textbf{mclust} R package (Fraley, Raftery, and
Scrucca 2023) used throughout this chapter, using tidyverse syntax and
terminology which is more familiar in the LPA literature.

\hypertarget{finite-gaussian-mixture-models}{%
\subsection{Finite Gaussian mixture
models}\label{finite-gaussian-mixture-models}}

As described above, finite mixture models (FMMs) provide the statistical
framework for model-based clustering and allow for the modelling of
complex data by combining simpler distributions. Specifically, a FMM
assumes that the observed data are generated from a finite mixture of
underlying distributions, each of which corresponds to a distinct
subgroup or cluster within the data. Gaussian mixture models (GMMs) are
a particularly widespread variant of FMMs which specifically assume that
each of the underlying distributions is a (multivariate) Gaussian
distribution. This means that the data within each cluster are normally
distributed, but with potentially different means and covariance
matrices. In this type of model, the latent variable represents the
cluster assignment for each observation in the data; it is an indicator
variable that takes the value 1 for the cluster to which the observation
belongs and 0 for all the other clusters. It is well-known that any
continuous density can be well-fitted by mixtures of Gaussians to
arbitrary accuracy.

To estimate the parameters of a GMM with the associated latent variable
for cluster membership, a likelihood-based approach is typically used.
The likelihood function expresses the probability of observing the data,
given the parameter values and the latent variable. The maximum
likelihood estimation (MLE) method is commonly used to estimate the
parameters and the latent variable which maximize the likelihood
function. Usually, the number of clusters in a GMM is also unknown, and
it is determined through a process known as model selection, which
involves comparing models with different numbers of clusters and
parameterisations and selecting the one which best fits the data.

In summary, model-based clustering from the perspective of latent
variable modelling assumes that the data is generated from a
probabilistic model with a specific number of clusters. A
likelihood-based approach can be used to estimate the parameters of the
model and the latent variable that represents the cluster assignment for
each observation in the data, and guide the selection of the number of
clusters. A GMM is a common framework for model-based clustering which
assumes the data in each cluster is generated from a Gaussian
distribution.

\hypertarget{gaussian-parsimonious-clustering-models}{%
\section{Gaussian parsimonious clustering
models}\label{gaussian-parsimonious-clustering-models}}

For a continuous feature vector \(\boldsymbol{x}\in \mathbb{R}^{d}\),
the general form of the density function of a Gaussian mixture model
(GMM) with \(K\) components can be written as
\begin{equation}\protect\hypertarget{eq-gmm}{}{
f(\boldsymbol{x}) = \sum_{k=1}^K \pi_k \phi_d(\boldsymbol{x};\boldsymbol{\mu}_k,\boldsymbol{\Sigma}_k),
}\label{eq-gmm}\end{equation} where \(\pi_k\) represents the mixing
probabilities which control the \textbf{size} of each cluster, such that
\(\pi_k > 0\) and \(\sum_{k=1}^K\pi_k=1\), and \(\phi_d(\cdot)\) is the
\(d\)-dimensional multivariate Gaussian density with parameters
\((\boldsymbol{\mu}_k,\boldsymbol{\Sigma}_k)\) for \(k=1,\ldots,K\).
Clusters described by such a GMM are ellipsoidal, centered at the means
\(\boldsymbol{\mu}_k\), and with other geometric characteristics (namely
volume, shape, and orientation) determined by the covariance matrices
\(\boldsymbol{\Sigma}_1, \ldots, \boldsymbol{\Sigma}_K\). Parsimonious
parameterisations of covariance matrices can be controlled by imposing
some constraints on the covariance matrices through the following
eigen-decomposition (Banfield and Raftery 1993; Celeux and Govaert
1995): \begin{equation}\protect\hypertarget{eq-eigendecomp}{}{
\boldsymbol{\Sigma}_k = \lambda_k \boldsymbol{U}_k \boldsymbol{\Delta}_k \boldsymbol{U}{}^{\!\top}_k,
}\label{eq-eigendecomp}\end{equation} where
\(\lambda_k = \lvert\boldsymbol{\Sigma}_k\rvert^{1/d}\) is a scalar
which controls the \emph{volume}, \(\boldsymbol{\Delta}_k\) is a
diagonal matrix whose entries are the normalized eigenvalues of
\(\boldsymbol{\Sigma}_k\) in decreasing order, such that
\(\lvert\boldsymbol{\Delta}_k\rvert = 1\), which controls the
\emph{shape} of the density contours, and \(\boldsymbol{U}_k\) is an
orthogonal matrix whose columns are the eigenvectors of
\(\boldsymbol{\Sigma}_k\), which controls the \emph{orientation} of the
corresponding ellipsoid.

Constraining these various quantities to be equal across clusters can
greatly reduce the number of estimable parameters, and is the means by
which GMMs obtain intermediate covariance matrices between
homoscedasticity and heteroscedasticity. A list of the 14 resulting
parameterisations available in the \textbf{mclust} package (Fraley,
Raftery, and Scrucca 2023) for R (R Core Team 2023) is included in Table
2.1 of Scrucca et al. (2023). Of particular note is the nomenclature
adopted by \textbf{mclust} whereby each model has a three-letter name
with each letter pertaining to the volume, shape, and orientation,
respectively, denoting whether the given component is equal (E) or free
to vary (V) across clusters. Some model names also use the letter I in
the third position to indicate that the covariance matrices are diagonal
and two particularly parsimonious models have the letter I in the second
position to indicate that the covariance matrices are isotropic. Thus,
as examples, the fully unconstrained VVV model is one for which the
volume, shape, and orientation are all free to vary across clusters, the
EVE model constrains the clusters to have equal volume and orientation
but varying shape, and the VII model assumes isotropic covariance
matrices with cluster-specific volumes. The flexibility to model
clusters with different geometric characteristics by modelling
correlations according to various parameterisations represents another
advantage over heuristic clustering algorithms. Taking the \(k\)-means
algorithm as an example, a larger number of circular, Euclidean
distance-based clusters may be required to fit the data well, rather
than a more parsimonious and easily interpretable mixture model with
fewer non-spherical components.

Given a random sample of observations
\(\{ \boldsymbol{x}_1, \boldsymbol{x}_2, \ldots, \boldsymbol{x}_n \}\)
in \(d\) dimensions, the log-likelihood of a GMM with \(K\) components
is given by \begin{equation}\protect\hypertarget{eq-loglik}{}{
\ell(\boldsymbol{\theta}) = \sum_{i=1}^n \log\left\{ \sum_{k=1}^K \pi_k \phi_d(\boldsymbol{x}_i ; \boldsymbol{\mu}_k, \boldsymbol{\Sigma}_k) \right\},
}\label{eq-loglik}\end{equation} where
\(\boldsymbol{\theta}= (\pi_1, \ldots, \pi_{K-1}, \boldsymbol{\mu}_1, \ldots, \boldsymbol{\mu}_K, \boldsymbol{\Sigma}_1, \ldots, \boldsymbol{\Sigma}_K)\)
denotes the collection of parameters to be estimated.

Maximizing the log-likelihood function in Equation~\ref{eq-loglik}
directly is often complicated, so maximum likelihood estimation (MLE) of
\(\boldsymbol{\theta}\) is usually performed using the EM algorithm
(Dempster, Laird, and Rubin 1977) by including the component membership
as a latent variable. The EM algorithm consists of two steps: the E-step
(Expectation step) and the M-step (Maximisation step). In the E-step,
the algorithm calculates the expected membership probabilities of each
data point to each of the mixture components based on the current
estimates of the model parameters. In the M-step, the algorithm updates
the model parameters by maximizing the likelihood of the observed data
given the estimated membership probabilities. These two steps are
repeated until convergence or a maximum number of iterations is reached.
Details on the use of EM algorithm in finite mixture modelling is
provided by McLachlan and Peel (2000), while a thorough treatment and
further extensions can be found in McLachlan and Krishnan (2008). For
the GMM case, see Sec. 2.2 of Scrucca et al. (2023).

Following the fitting of a GMM and the determination of the MLEs of
parameters, the maximum a posteriori (MAP) procedure can be used to
classify the observations into the most likely cluster and recover a
``hard'' partition. For an observation \(\boldsymbol{x}_i\) the
posterior conditional probability of coming from the mixture component
\(k\) is given by
\begin{equation}\protect\hypertarget{eq-postcondprob}{}{
\widehat{p}_{ik} = \frac{\widehat{\pi}_k \phi_d(\boldsymbol{x}_i; \widehat{\boldsymbol{\mu}}_k, \widehat{\boldsymbol{\Sigma}}_k)}{\displaystyle\sum_{g=1}^K \widehat{\pi}_g \phi_d(\boldsymbol{x}; \widehat{\boldsymbol{\mu}}_g, \widehat{\boldsymbol{\Sigma}}_g)}.
}\label{eq-postcondprob}\end{equation} Then, an observation is assigned
to the cluster with the largest posterior conditional probability, i.e.,
\(\boldsymbol{x}_i \in \mathcal{C}_{k^\star}\) with
\(k^\star = \mathop{\mathrm{arg\,max}}_k \widehat{p}_{ik}\).

\hypertarget{model-selection}{%
\subsection{Model selection}\label{model-selection}}

Given that a wide variety of GMMs in Equation~\ref{eq-gmm} can be
estimated by varying the number of mixture components and the covariance
decompositions in Equation~\ref{eq-eigendecomp}, selecting the
appropriate model represents a crucial decision. A popular option
consists in choosing the ``best'' model using the Bayesian information
criterion (BIC, Schwarz 1978), which, for a given model \(\mathcal{M}\),
is defined as \[
\text{BIC}_{\mathcal{M}} = 2\ell_{\mathcal{M}}(\widehat{\boldsymbol{\theta}}) - \nu_{\mathcal{M}} \log(n),
\] where \(\ell_{\mathcal{M}}(\widehat{\boldsymbol{\theta}})\) stands
for the maximized log-likelihood of the data sample of size \(n\) under
model \(\mathcal{M}\), and \(\nu_{\mathcal{M}}\) for the number of
independent parameters to be estimated. Another option available in
clustering is the Integrated Complete Likelihood (ICL, Biernacki,
Celeux, and Govaert 2000) criterion given by \[
\text{ICL}_{\mathcal{M}} = \text{BIC}_{\mathcal{M}} + 2 \sum_{i=1}^n\sum_{k=1}^K c_{ik} \log(\widehat{p}_{ik}),
\] where \(\widehat{p}_{ik}\) is the conditional probability that
\(\boldsymbol{x}_i\) arises from the \(k\)-th mixture component from
Equation~\ref{eq-postcondprob}, and \(c_{ik} = 1\) if the \(i\)-th
observation is assigned to cluster \(\mathcal{C}_k\) and 0 otherwise.

Both criteria evaluate the fit of a GMM to a given set of data by
considering both the likelihood of the data given the model and the
complexity of the model itself, represented by the number of parameters
to be estimated. Compared to the BIC, the ICL introduces a further
entropy-based penalisation for the overlap of the clusters. For this
reason, the ICL tends to select models with well-separated clusters.

Whereas there is no consensus of a standard criteria for choosing the
best model, there are guidelines that the researcher could rely on. To
decide on the optimal model, examining the fit indices (such as the BIC
and ICL), model interpretability, and conformance to theory can be of
great help. The literature recommends estimating a 1-class solution for
each model that serves as a comparative baseline and then increasing the
number of classes one by one evaluating if adding another class yields a
better solution in both statistical and conceptual terms (Nylund-Gibson
and Choi 2018). Among all fit indices, lower BIC values seems to be the
preferred method for selecting the best model. However, examining other
indices (e.g., AIC, ICL) is also useful. Oftentimes, fit indices do not
converge to a certain model. In such cases, the interrelation between
the selected models, such as whether one model is an expanded version of
another, should also be taken into consideration, as well as the
stability of the different models, including the relative sizes of the
emergent classes (each class should comprise more than 5-8\% of the
sample) (Nylund-Gibson and Choi 2018). Furthermore, the elbow method
could be helpful in cases where no clear number of clusters can be
easily determined from the fit indices (e.g., the BIC continues to
decrease consistently when increasing the number of classes). This
entails plotting the BIC values and finding an elbow shape where a drop
in BIC is less noticeable with increasing number of classes or roughly
an elbow followed by a relatively flat line. The choice of the best
number of classes can and probably should be guided by theory; that is,
in cases where previous research reported a certain number of clusters
or patterns, it is recommended to take this guidance into account. For
instance, research on engagement has repeatedly reported three levels of
engagement. Once we have chosen the most suitable model, it is suggested
to compute model diagnostics (e.g., entropy and average posterior
probability) to evaluate the selected model. These diagnostics are
covered in Section~\ref{sec-entropy}.

\hypertarget{mclust-r-package}{%
\subsection{mclust R package}\label{mclust-r-package}}

\textbf{mclust} is an R package (R Core Team 2023) for model-based
cluster analysis, classification, and density estimation using Gaussian
finite mixture models (Scrucca et al. 2016, 2023). It is widely used in
statistics, machine learning, data science, and pattern recognition. One
of the key features of \textbf{mclust} is its flexibility in modelling
quantitative data with several covariance structures and different
numbers of mixture components. Additionally, the package provides
extensive graphical representations, model selection criteria,
initialisation strategies for the EM algorithm, bootstrap-based
inference, and Bayesian regularisation, among other prominent features.
\textbf{mclust} also represents a valuable tool in educational settings
because it provides a powerful set of models that allows students and
researchers to quickly and easily perform clustering and classification
analyses on their data.

The main function implementing model-based clustering is called
\texttt{Mclust()}, which requires a user to provide at least the data
set to analyze. In the one-dimensional case, the data set can be a
vector, while in the multivariate case, it can be a matrix or a data
frame. In the latter case, the rows correspond to observations, and the
columns correspond to variables.

The \texttt{Mclust()} function allows for further arguments, including
the optional argument \texttt{G} to specify the number of mixture
components or clusters, and \texttt{modelNames} to specify the
covariance decomposition. If both \texttt{G} and \texttt{modelNames} are
not provided, \texttt{Mclust()} will fit all possible models obtained
using a number of mixture components from 1 to 9 and all 14 available
covariance decompositions, and it will select the model with the largest
BIC. Notably, if the data set is univariate, only 2 rather than 14
models governing the scalar variance parameters are returned; that they
are equal or unequal across components. Finally, computing the BIC and
ICL criteria can be done by invoking the functions \texttt{mclustBIC()}
and \texttt{mclustICL()}, respectively.

\hypertarget{other-practical-issues-and-extensions}{%
\subsection{Other practical issues and
extensions}\label{other-practical-issues-and-extensions}}

Prior to commencing the cluster analysis of a data set on school
engagement, academic achievement, and self-regulated learning measures,
we first provide some theoretical background on some extensions of
practical interest which will be explored in the analysis.

\hypertarget{bayesian-regularisation}{%
\subsubsection{Bayesian regularisation}\label{bayesian-regularisation}}

Including a prior distribution over the mixture parameters is an
effective way to avoid singularities and degeneracies in maximum
likelihood estimation. Furthermore, this can help to prevent overfitting
and improve model performance. In latent class analysis, where the
variables of interest are often discrete or take on only a few values,
including a prior distribution can help to regularize the model.

Fraley and Raftery (2007) proposed using weekly informative conjugate
priors to regularize the estimation process. The EM algorithm can still
be used for model fitting, but maximum likelihood estimates (MLEs) are
replaced by maximum a posteriori (MAP) estimates. A slightly modified
version of BIC can be used for model selection, with the maximized
log-likelihood replaced by the log-likelihood evaluated at the MAP or
posterior mode.

The prior distributions proposed by Fraley and Raftery (2007) are:

\begin{itemize}
\item
  a uniform prior on the simplex for the mixture weights
  \((\pi_1, \ldots, \pi_K)\);
\item
  a Gaussian prior on the mean vector (conditional on the covariance
  matrix), i.e., \begin{align}
  \boldsymbol{\mu}\mid \boldsymbol{\Sigma}
  & \sim \mathcal{N}(\boldsymbol{\mu}_{\scriptscriptstyle \mathcal{P}}, \boldsymbol{\Sigma}/\kappa_{\scriptscriptstyle \mathcal{P}}) \\
  & \propto
  \left|\boldsymbol{\Sigma}\right|^{-1/2}
  \exp\left\{ -\frac{\kappa_{\scriptscriptstyle \mathcal{P}}}{2}
              \mathop{\mathrm{tr}}\left(\left(\boldsymbol{\mu}- \boldsymbol{\mu}_{\scriptscriptstyle \mathcal{P}}\right){}^{\!\top}
                       \boldsymbol{\Sigma}^{-1} 
                       \left(\boldsymbol{\mu}- \boldsymbol{\mu}_{\scriptscriptstyle \mathcal{P}}\right)
                  \right) 
      \right\},
  \label{eqn:multivariatePriorMean}
  \end{align} with \(\boldsymbol{\mu}_{\scriptscriptstyle \mathcal{P}}\)
  and \(\kappa_{\scriptscriptstyle \mathcal{P}}\) being the
  hyperparameters controlling, respectively, the mean vector and the
  amount of shrinkage applied;
\item
  an inverse Wishart prior on the covariance matrix, i.e., \begin{align}
  \boldsymbol{\Sigma}
  & \sim \mathcal{IW}(\nu_{\scriptscriptstyle \mathcal{P}}, \boldsymbol{\Lambda}_{\scriptscriptstyle \mathcal{P}})
  \nonumber\\
  & \propto
  \left|\boldsymbol{\Sigma}\right|^{-(\nu_{\scriptscriptstyle \mathcal{P}}+d+1)/2}
  \exp\left\{ -\frac{1}{2}
              \mathop{\mathrm{tr}}\left(\boldsymbol{\Sigma}^{-1} \boldsymbol{\Lambda}_{\scriptscriptstyle \mathcal{P}}^{-1}
                 \right)
      \right\},
  \label{eqn:multivariatePriorVar}
  \end{align} with the hyperparameters
  \(\nu_{\scriptscriptstyle \mathcal{P}}\) and the matrix
  \(\boldsymbol{\Lambda}_{\scriptscriptstyle \mathcal{P}}\) controlling
  the degrees of freedom and scale of the prior distribution,
  respectively.
\end{itemize}

Adding a prior to GMMs estimated using the \textbf{mclust} R package is
easily obtained by adding an optional \texttt{prior} argument when
calling some of the fitting functions, such as \texttt{mclustBIC()} and
\texttt{Mclust()}. Specifically, setting
\texttt{prior\ =\ priorControl(functionName\ =\ "defaultPrior")} allows
to adopt the conjugate priors described above with the following default
values for the hyperparameters:

\begin{itemize}
\item
  mean vector
  \(\boldsymbol{\mu}_{\scriptscriptstyle \mathcal{P}}= \bar{\boldsymbol{x}}\),
  the sample mean of each variable;
\item
  shrinkage \(\kappa_{\scriptscriptstyle \mathcal{P}}= 0.1\);
\item
  degrees of freedom \(\nu_{\scriptscriptstyle \mathcal{P}}= d+2\);
\item
  scale matrix
  \(\boldsymbol{\Lambda}_{\scriptscriptstyle \mathcal{P}}= \S/(K^{2/d})\),
  where \(\S\) is the sample covariance matrix.
\end{itemize}

Rationale for the above default values for the prior hyperparameters,
together with the corresponding MAP estimates of the GMM parameters, can
be found in Fraley and Raftery (2007, Table 2). These values should
suffice for most applications, but experienced users who want to tune
the hyperparameters can refer to the documentation available in the help
pages for \texttt{priorControl} and \texttt{defaultPrior()}. Further
details about specifying different hyperparameter values can be found in
Scrucca et al. (2023, sec. 7.2).

\hypertarget{bootstrap-inference}{%
\subsubsection{Bootstrap inference}\label{bootstrap-inference}}

Likelihood-based inference in mixture models is complicated because
asymptotic theory applied to mixture models require a very large sample
size (McLachlan and Peel 2000, 299:42), and standard errors derived from
the expected or the observed information matrix tend to be unstable
(Basford et al. 1997). For these reasons, resampling approaches based on
the bootstrap are often employed (O'Hagan et al. 2019).

The \emph{bootstrap} (Efron 1979) is a general, widely applicable,
powerful technique for obtaining an approximation to the sampling
distribution of a statistic of interest. The bootstrap distribution is
approximated by drawing a large number of samples, called
\emph{bootstrap samples}, from the empirical distribution. This can be
obtained by resampling with replacement from the observed data
(\emph{nonparametric bootstrap}), or from a parametric distribution with
unknown parameters substituted by the corresponding estimates
(\emph{parametric bootstrap}). A Bayesian version of the bootstrap,
introduced by Rubin (1981), allows posterior samples to be obtained by
resampling with weights for each observation drawn from a uniform
Dirichlet distribution. A strictly related technique is the
\emph{weighted likelihood bootstrap} (Newton and Raftery 1994), where a
statistical model is repeatedly fitted using weighted maximum likelihood
with weights obtained as in Bayesian bootstrap.

Let \(\widehat{\boldsymbol{\theta}}\) be the estimate of a set of GMM
parameters \(\boldsymbol{\theta}\) for a given model \(\mathcal{M}\),
determined by the adopted covariance parameterisation and number of
mixture components. The bootstrap distribution for the parameters of
interest is obtained as follows:

\begin{itemize}
\item
  draw a bootstrap sample of size \(n\) using one of the resampling
  techniques described above to form the bootstrap sample
  \((\boldsymbol{x}^\star_1, \ldots, \boldsymbol{x}^\star_n)\);
\item
  fit a the GMM \(\mathcal{M}\) to get the bootstrap estimates
  \(\widehat{\boldsymbol{\theta}}^\star\);
\item
  replicate the previous steps a large number of times, say \(B\).
\end{itemize}

The bootstrap distribution for the parameters of interest,
\(\widehat{\boldsymbol{\theta}}^\star_1, \widehat{\boldsymbol{\theta}}^\star_2, \ldots, \widehat{\boldsymbol{\theta}}^\star_B\),
can then be used for computing the bootstrap standard errors (as the
square root of the diagonal elements of the bootstrap covariance matrix)
or the bootstrap percentile confidence intervals. More details can be
found in Scrucca et al. (2023, sec. 2.4).

From a practical point of view, bootstrap resampling can be conducted in
\textbf{mclust} by means of the function \texttt{MclustBootstrap()}.
This function takes as arguments the fitted model object returned from
e.g., \texttt{Mclust()} or \texttt{mclustBIC()}, the optional argument
\texttt{type}, which allows to specify the type of bootstrap samples to
draw (\texttt{"bs"} for nonparametric bootstrap, \texttt{"pb"} for
parametric bootstrap, and \texttt{"wlbs"} for weighted likelihood
bootstrap), and the optional argument \texttt{nboot}, which sets the
number of bootstrap samples. At least \(999\) samples should be drawn if
confidence intervals are needed.

\hypertarget{sec-entropy}{%
\subsubsection{Entropy and average posterior
probabilities}\label{sec-entropy}}

The definition of entropy in information theory (Cover and Thomas 2006)
refers to the average amount of information provided by a random
variable. Following this definition, Celeux and Soromenho (1996) defines
the entropy of a finite mixture model as follows \[
E_{\text{FMM}} = - \sum_{i=1}^n \sum_{k=1}^K \widehat{p}_{ik} \log(\widehat{p}_{ik}),
\] where \(\widehat{p}_{ik}\) is the estimated posterior probability of
case \(i\) to belong to cluster \(k\) (see
Equation~\ref{eq-postcondprob}). If the mixture components are well
separated, \(\widehat{p}_{ik} \approx 1\) for the assigned cluster
\(\mathcal{C}_k\) and 0 otherwise. Consequently, the entropy of the
mixture model in this case is \(E_{\text{FMM}} = 0\) (note that
\(0\log(0)=0\) by convention). On the contrary, in the case of maximal
classification uncertainty, \(\widehat{p}_{ik} = 1/K\) for all clusters
\(\mathcal{C}_k\) (\(k=1,\ldots,K\)). As a result, the entropy of the
mixture model is \(E_{\text{FMM}} = n\log(K)\).

In latent class and latent profile analysis, a slightly different
definition of entropy is used as a diagnostic statistic to assess how
well the fitted model assigns individuals to the identified clusters
based on their response patterns. Thus, a normalized version of the
entropy is defined as follows \[
E = 1 - \frac{E_{\text{FMM}}}{n \log(K)} = 1 + \dfrac{\sum_{i=1}^n \sum_{k=1}^K \widehat{p}_{ik} \log(\widehat{p}_{ik})}{n \log(K)}.
\] Entropy takes values on the range \([0,1]\), with higher entropy
values indicating that the model has less uncertainty in assigning cases
to their respective latent classes/profiles. Thus, high entropy values
typically indicate a better model which is able to distinguish between
the latent components and that the components are relatively distinct.
An entropy value close to 1 is ideal, while values above \(0.6\) are
considered acceptable, although there is no agreed upon optimal cutoff
for entropy.

The contribution of each observation to the overall entropy can be
defined as \[
E_i = 1 + \frac{\sum_{k=1}^K \widehat{p}_{ik} \log(\widehat{p}_{ik})}{\log(K)},
\] so that the overall entropy is obtained by averaging over the single
contributions, i.e., \(E = \sum_{i=1}^n E_i/n\). The single
contributions \(E_i\) can also be used to compute the average entropy of
each latent component, which indicates how accurately the model defines
components. Average posterior probabilities (AvePP) are a closely
related performance assessment measure, given by the average posterior
membership probabilities \(\widehat{p}_{ik}\) for each component for the
observations most probably assigned to that component, for which a
cutoff of \(0.8\) has been suggested to indicate acceptably high
assignment certainty and well-separated clusters (Nylund-Gibson and Choi
2018). The analysis below presents the necessary code to calculate
entropies and average posterior probabilities thusly from a fitted
\textbf{mclust} model.

\hypertarget{application-school-engagement-academic-achievement-and-self-regulated-learning}{%
\section{Application: School engagement, academic achievement, and
self-regulated
learning}\label{application-school-engagement-academic-achievement-and-self-regulated-learning}}

A group of primary school students from northern Spain were evaluated in
terms of their school engagement, self-regulation, and academic
performance through the use of various measures. The school engagement
measure (SEM) was employed to assess their engagement, while their
self-regulation was evaluated with the self-regulation strategy
inventory---self-report. The measure for academic achievement was based
on the students' self-reported grades in Spanish and mathematics, which
were rated on a scale of 1 to 5. This data set can be used to identify
clusters of students based on their engagement and self-regulation.
These clusters would represent distinct patterns or ``profiles'' of
engagement. Finding such profiles would allow us to understand
individual differences but more importantly to stratify support
according to different engagement profiles.

\hypertarget{preparing-the-data}{%
\subsection{Preparing the data}\label{preparing-the-data}}

We start by loading the packages required for the analysis. We note in
particular that version 6.0.0 of \textbf{mclust} is employed here, the
latest release at the time of writing.

\begin{Shaded}
\begin{Highlighting}[]
\FunctionTok{library}\NormalTok{(ggplot2)}
\FunctionTok{library}\NormalTok{(ggridges)}
\FunctionTok{library}\NormalTok{(mclust)}
\FunctionTok{library}\NormalTok{(rio)}
\FunctionTok{library}\NormalTok{(tidyverse)}
\end{Highlighting}
\end{Shaded}

Then, we read the data set from an online comma-separated-value (CSV)
file, followed by some data cleaning and formatting to prepare the data
for subsequent analysis. Note that the CSV file to be read is not in
standard format, so we have to explicitly set the separator field using
the optional argument \texttt{sep\ =\ ";"}.

\tiny

\begin{Shaded}
\begin{Highlighting}[]
\CommentTok{\# read the data}
\NormalTok{data }\OtherTok{\textless{}{-}} \FunctionTok{import}\NormalTok{(}\StringTok{"https://raw.githubusercontent.com/sonsoleslp/labook{-}data/main/3\_engSRLach/Manuscript\_School\%20Engagment.csv"}\NormalTok{, }\AttributeTok{sep =} \StringTok{";"}\NormalTok{)}
\end{Highlighting}
\end{Shaded}

\normalsize

\begin{Shaded}
\begin{Highlighting}[]
\CommentTok{\# select the variables to be analyzed}
\NormalTok{vars }\OtherTok{\textless{}{-}} \FunctionTok{c}\NormalTok{(}\StringTok{"PRE\_ENG\_COND"}\NormalTok{, }\StringTok{"PRE\_ENG\_COGN"}\NormalTok{, }\StringTok{"PRE\_ENG\_EMOC"}\NormalTok{)}
\NormalTok{x }\OtherTok{\textless{}{-}} \FunctionTok{select}\NormalTok{(data, }\FunctionTok{all\_of}\NormalTok{(vars)) }\SpecialCharTok{|\textgreater{}} 
  \FunctionTok{as\_tibble}\NormalTok{() }\SpecialCharTok{|\textgreater{}}
  \FunctionTok{rename}\NormalTok{(}\StringTok{"BehvEngmnt"} \OtherTok{=} \StringTok{"PRE\_ENG\_COND"}\NormalTok{,  }\CommentTok{\# Behavioral engagement}
         \StringTok{"CognEngmnt"} \OtherTok{=} \StringTok{"PRE\_ENG\_COGN"}\NormalTok{,  }\CommentTok{\# Cognitive engagement}
         \StringTok{"EmotEngmnt"} \OtherTok{=} \StringTok{"PRE\_ENG\_EMOC"}\NormalTok{)  }\CommentTok{\# Emotional engagement}
\CommentTok{\# print the data set used in the subsequent analysis}
\NormalTok{x}
\DocumentationTok{\#\# \# A tibble: 717 x 3}
\DocumentationTok{\#\#    BehvEngmnt CognEngmnt EmotEngmnt}
\DocumentationTok{\#\#         \textless{}dbl\textgreater{}      \textless{}dbl\textgreater{}      \textless{}dbl\textgreater{}}
\DocumentationTok{\#\#  1       3.75       3.14        4.4}
\DocumentationTok{\#\#  2       4          3.71        2  }
\DocumentationTok{\#\#  3       4.25       3.86        4  }
\DocumentationTok{\#\#  4       3.75       2.57        3  }
\DocumentationTok{\#\#  5       4.25       3           4  }
\DocumentationTok{\#\#  6       4          3.71        3.8}
\DocumentationTok{\#\#  7       3.5        2.14        3.2}
\DocumentationTok{\#\#  8       4.75       3.57        1.6}
\DocumentationTok{\#\#  9       3.25       2.71        3  }
\DocumentationTok{\#\# 10       5          4.43        4.8}
\DocumentationTok{\#\# \# i 707 more rows}
\end{Highlighting}
\end{Shaded}

A table of summary statistics for the data set can be obtained as
follows:

\begin{Shaded}
\begin{Highlighting}[]
\NormalTok{x }\SpecialCharTok{|\textgreater{}} \FunctionTok{pivot\_longer}\NormalTok{(}\AttributeTok{cols =} \FunctionTok{colnames}\NormalTok{(x)) }\SpecialCharTok{|\textgreater{}}
  \FunctionTok{group\_by}\NormalTok{(name) }\SpecialCharTok{|\textgreater{}}
  \FunctionTok{summarize}\NormalTok{(}\AttributeTok{N =} \FunctionTok{n}\NormalTok{(),}
            \AttributeTok{Nunq =} \FunctionTok{n\_distinct}\NormalTok{(value),}
            \AttributeTok{Mean =} \FunctionTok{mean}\NormalTok{(value),}
            \AttributeTok{SD =} \FunctionTok{sd}\NormalTok{(value),}
            \AttributeTok{Min =} \FunctionTok{min}\NormalTok{(value),}
            \AttributeTok{Median =} \FunctionTok{median}\NormalTok{(value),}
            \AttributeTok{Max =} \FunctionTok{max}\NormalTok{(value))}
\DocumentationTok{\#\# \# A tibble: 3 x 8}
\DocumentationTok{\#\#   name           N  Nunq  Mean    SD   Min Median   Max}
\DocumentationTok{\#\#   \textless{}chr\textgreater{}      \textless{}int\textgreater{} \textless{}int\textgreater{} \textless{}dbl\textgreater{} \textless{}dbl\textgreater{} \textless{}dbl\textgreater{}  \textless{}dbl\textgreater{} \textless{}dbl\textgreater{}}
\DocumentationTok{\#\# 1 BehvEngmnt   717    17  4.17 0.627     1   4.25     5}
\DocumentationTok{\#\# 2 CognEngmnt   717    30  2.92 0.771     1   2.92     5}
\DocumentationTok{\#\# 3 EmotEngmnt   717    22  3.61 0.911     1   3.61     5}
\end{Highlighting}
\end{Shaded}

\hypertarget{model-estimation-and-model-selection}{%
\subsection{Model estimation and model
selection}\label{model-estimation-and-model-selection}}

To begin our latent profile analysis, we first fit a number of candidate
GMMs with different numbers of latent components and covariance
parameritations, and compute the Bayesian Information Criterion (BIC) to
select the ``optimal'' model. This model selection criterion jointly
takes into account both the covariance decompositions and the number of
mixture components in the model.

As mentioned earlier, given the characteristics of the data, which
consists of a small number of unique values relative to the number of
observations, a prior is used for regularisation. We invoke the default
priors described above, summarise the BIC values of the three best
models, and visualise the BIC values of all fitted models.

\begin{Shaded}
\begin{Highlighting}[]
\NormalTok{BIC }\OtherTok{\textless{}{-}} \FunctionTok{mclustBIC}\NormalTok{(x, }\AttributeTok{prior =} \FunctionTok{priorControl}\NormalTok{())}
\FunctionTok{summary}\NormalTok{(BIC)}
\DocumentationTok{\#\# Best BIC values:}
\DocumentationTok{\#\#              VVI,3        VVI,4       VVV,3}
\DocumentationTok{\#\# BIC      {-}4521.213 {-}4526.905884 {-}4533.57166}
\DocumentationTok{\#\# BIC diff     0.000    {-}5.693183   {-}12.35896}
\FunctionTok{plot}\NormalTok{(BIC)}
\end{Highlighting}
\end{Shaded}

\begin{figure}[H]

{\centering \includegraphics{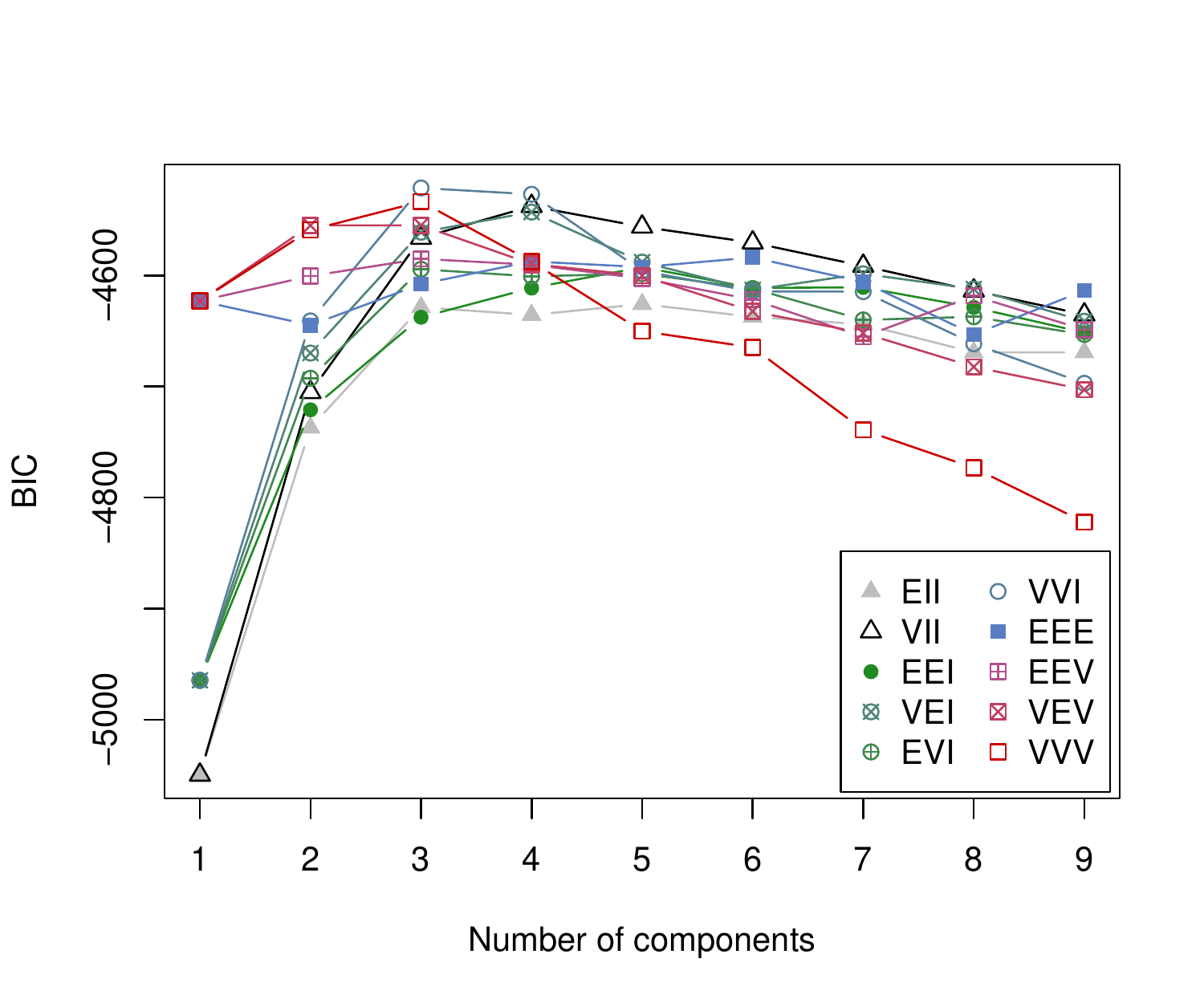}

}

\end{figure}

The selected model is a three-component GMM with diagonal covariance
matrices of varying volume and shape, with axis-aligned orientation,
indicated as `(VVI,3)'. Thus, the variables are independent within each
cluster.

\hypertarget{examining-model-output}{%
\subsection{Examining model output}\label{examining-model-output}}

The fit of the optimal model is obtained using:

\begin{Shaded}
\begin{Highlighting}[]
\NormalTok{mod }\OtherTok{\textless{}{-}} \FunctionTok{Mclust}\NormalTok{(x, }\AttributeTok{modelNames =} \StringTok{"VVI"}\NormalTok{, }\AttributeTok{G =} \DecValTok{3}\NormalTok{, }\AttributeTok{prior =} \FunctionTok{priorControl}\NormalTok{())}
\FunctionTok{summary}\NormalTok{(mod, }\AttributeTok{parameters =} \ConstantTok{TRUE}\NormalTok{)}
\DocumentationTok{\#\# {-}{-}{-}{-}{-}{-}{-}{-}{-}{-}{-}{-}{-}{-}{-}{-}{-}{-}{-}{-}{-}{-}{-}{-}{-}{-}{-}{-}{-}{-}{-}{-}{-}{-}{-}{-}{-}{-}{-}{-}{-}{-}{-}{-}{-}{-}{-}{-}{-}{-}{-}{-} }
\DocumentationTok{\#\# Gaussian finite mixture model fitted by EM algorithm }
\DocumentationTok{\#\# {-}{-}{-}{-}{-}{-}{-}{-}{-}{-}{-}{-}{-}{-}{-}{-}{-}{-}{-}{-}{-}{-}{-}{-}{-}{-}{-}{-}{-}{-}{-}{-}{-}{-}{-}{-}{-}{-}{-}{-}{-}{-}{-}{-}{-}{-}{-}{-}{-}{-}{-}{-} }
\DocumentationTok{\#\# }
\DocumentationTok{\#\# Mclust VVI (diagonal, varying volume and shape) model with 3 components: }
\DocumentationTok{\#\# }
\DocumentationTok{\#\# Prior: defaultPrior() }
\DocumentationTok{\#\# }
\DocumentationTok{\#\#  log{-}likelihood   n df       BIC       ICL}
\DocumentationTok{\#\#       {-}2194.856 717 20 {-}4521.213 {-}4769.174}
\DocumentationTok{\#\# }
\DocumentationTok{\#\# Clustering table:}
\DocumentationTok{\#\#   1   2   3 }
\DocumentationTok{\#\# 184 119 414 }
\DocumentationTok{\#\# }
\DocumentationTok{\#\# Mixing probabilities:}
\DocumentationTok{\#\#         1         2         3 }
\DocumentationTok{\#\# 0.2895147 0.1620776 0.5484078 }
\DocumentationTok{\#\# }
\DocumentationTok{\#\# Means:}
\DocumentationTok{\#\#                [,1]     [,2]     [,3]}
\DocumentationTok{\#\# BehvEngmnt 3.704041 4.713234 4.257355}
\DocumentationTok{\#\# CognEngmnt 2.287057 3.699530 3.017293}
\DocumentationTok{\#\# EmotEngmnt 2.738969 4.733899 3.737286}
\DocumentationTok{\#\# }
\DocumentationTok{\#\# Variances:}
\DocumentationTok{\#\# [,,1]}
\DocumentationTok{\#\#            BehvEngmnt CognEngmnt EmotEngmnt}
\DocumentationTok{\#\# BehvEngmnt  0.5022148  0.0000000  0.0000000}
\DocumentationTok{\#\# CognEngmnt  0.0000000  0.3909235  0.0000000}
\DocumentationTok{\#\# EmotEngmnt  0.0000000  0.0000000  0.7674268}
\DocumentationTok{\#\# [,,2]}
\DocumentationTok{\#\#            BehvEngmnt CognEngmnt EmotEngmnt}
\DocumentationTok{\#\# BehvEngmnt  0.0737948  0.0000000 0.00000000}
\DocumentationTok{\#\# CognEngmnt  0.0000000  0.4150514 0.00000000}
\DocumentationTok{\#\# EmotEngmnt  0.0000000  0.0000000 0.05540526}
\DocumentationTok{\#\# [,,3]}
\DocumentationTok{\#\#            BehvEngmnt CognEngmnt EmotEngmnt}
\DocumentationTok{\#\# BehvEngmnt  0.2048374  0.0000000  0.0000000}
\DocumentationTok{\#\# CognEngmnt  0.0000000  0.3327557  0.0000000}
\DocumentationTok{\#\# EmotEngmnt  0.0000000  0.0000000  0.2795838}
\end{Highlighting}
\end{Shaded}

The shown output reports some basic information about the fit, such as
the maximized log-likelihood (\texttt{log-likelihood}), the number of
observations (\texttt{n}), the number of estimated parameters
(\texttt{df}), the BIC criterion (\texttt{BIC}), and the clustering
table based on the MAP classification. The latter indicates that the
clusters also vary in terms of \emph{size}. The optional argument
\texttt{parameters\ =\ TRUE} in the \texttt{summary()} function call
additionally prints the estimated parameters. Observe that the VVI model
allows variance to vary across components while fixing all covariance
parameters to zero.

A plot showing the classification provided by the estimated model can be
drawn as follows:

\begin{Shaded}
\begin{Highlighting}[]
\FunctionTok{plot}\NormalTok{(mod, }\AttributeTok{what =} \StringTok{"classification"}\NormalTok{) }
\end{Highlighting}
\end{Shaded}

\begin{figure}[H]

{\centering \includegraphics{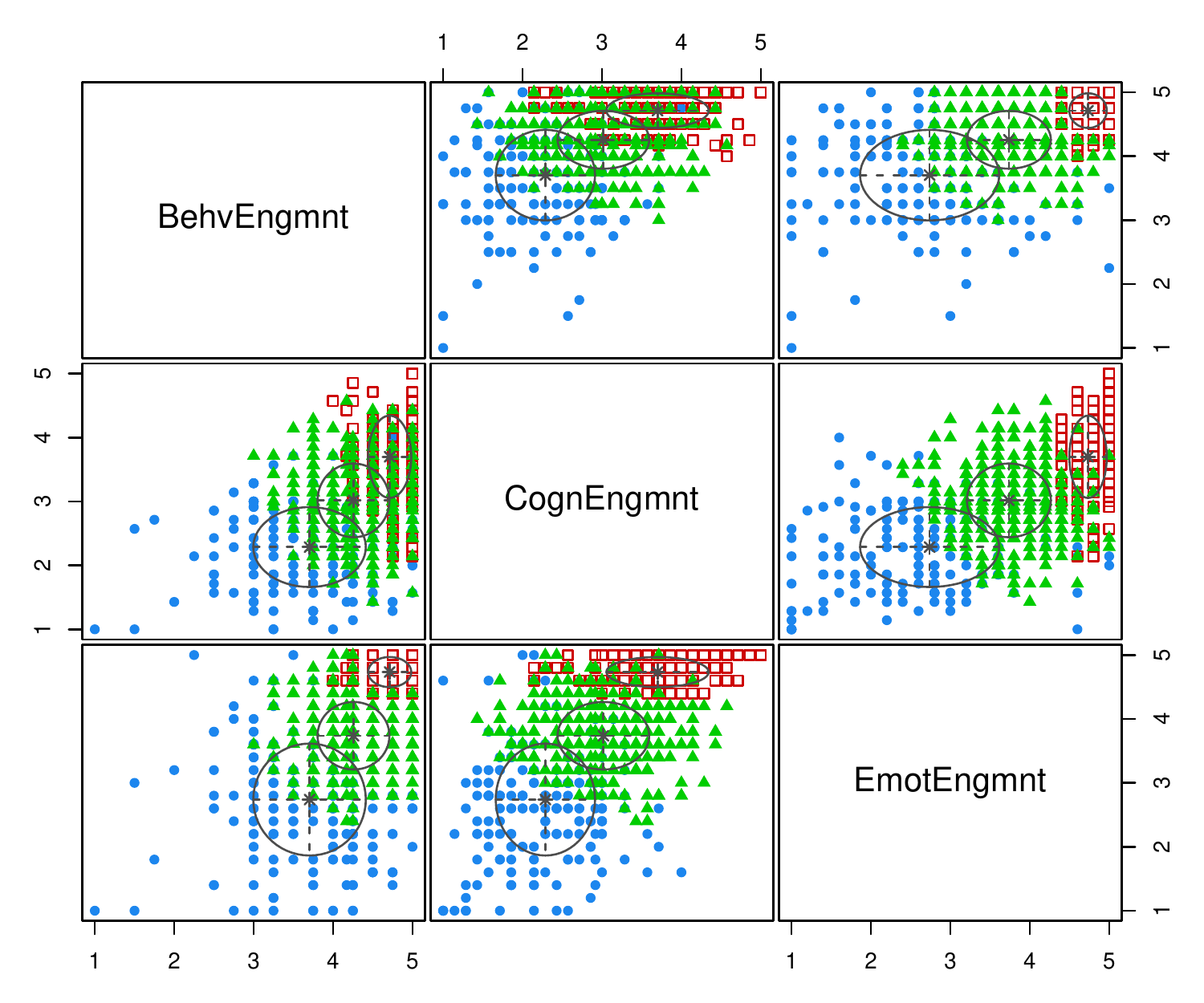}

}

\end{figure}

The estimated model identifies three clusters of varying size. The third
group (shown as filled green triangles) accounts for more than 50\% of
the observations, while the first (shown as blue filled points) and the
second (shown as red open squares) account for approximately 29\% and
16\%, respectively. The smallest cluster is also the group with the
largest engagement scores.

The different engagement behaviour of the three identified clusters can
be shown using a latent profiles plot of the estimated means with point
sizes proportional to the estimated mixing probabilities:

\begin{Shaded}
\begin{Highlighting}[]
\CommentTok{\# collect estimated means}
\NormalTok{means }\OtherTok{\textless{}{-}} \FunctionTok{data.frame}\NormalTok{(}\AttributeTok{Profile =} \FunctionTok{factor}\NormalTok{(}\DecValTok{1}\SpecialCharTok{:}\NormalTok{mod}\SpecialCharTok{$}\NormalTok{G),}
                    \FunctionTok{t}\NormalTok{(mod}\SpecialCharTok{$}\NormalTok{parameters}\SpecialCharTok{$}\NormalTok{mean)) }\SpecialCharTok{|\textgreater{}}
  \FunctionTok{pivot\_longer}\NormalTok{(}\AttributeTok{cols =} \SpecialCharTok{{-}}\DecValTok{1}\NormalTok{,}
               \AttributeTok{names\_to =} \StringTok{"Variable"}\NormalTok{,}
               \AttributeTok{values\_to =} \StringTok{"Mean"}\NormalTok{)}
\CommentTok{\# convert variable names to factor}
\NormalTok{means}\SpecialCharTok{$}\NormalTok{Variable }\OtherTok{\textless{}{-}} \FunctionTok{factor}\NormalTok{(means}\SpecialCharTok{$}\NormalTok{Variable, }
                         \AttributeTok{levels =} \FunctionTok{colnames}\NormalTok{(mod}\SpecialCharTok{$}\NormalTok{data))}
\CommentTok{\# add mixing probabilities corresponding to profiles}
\NormalTok{means }\OtherTok{\textless{}{-}}\NormalTok{ means }\SpecialCharTok{|\textgreater{}} 
  \FunctionTok{add\_column}\NormalTok{(}\AttributeTok{MixPro =}\NormalTok{ mod}\SpecialCharTok{$}\NormalTok{parameters}\SpecialCharTok{$}\NormalTok{pro[means}\SpecialCharTok{$}\NormalTok{Profile])}
\NormalTok{means}
\DocumentationTok{\#\# \# A tibble: 9 x 4}
\DocumentationTok{\#\#   Profile Variable    Mean MixPro}
\DocumentationTok{\#\#   \textless{}fct\textgreater{}   \textless{}fct\textgreater{}      \textless{}dbl\textgreater{}  \textless{}dbl\textgreater{}}
\DocumentationTok{\#\# 1 1       BehvEngmnt  3.70  0.290}
\DocumentationTok{\#\# 2 1       CognEngmnt  2.29  0.290}
\DocumentationTok{\#\# 3 1       EmotEngmnt  2.74  0.290}
\DocumentationTok{\#\# 4 2       BehvEngmnt  4.71  0.162}
\DocumentationTok{\#\# 5 2       CognEngmnt  3.70  0.162}
\DocumentationTok{\#\# 6 2       EmotEngmnt  4.73  0.162}
\DocumentationTok{\#\# 7 3       BehvEngmnt  4.26  0.548}
\DocumentationTok{\#\# 8 3       CognEngmnt  3.02  0.548}
\DocumentationTok{\#\# 9 3       EmotEngmnt  3.74  0.548}

\FunctionTok{ggplot}\NormalTok{(means, }\FunctionTok{aes}\NormalTok{(}\AttributeTok{x =}\NormalTok{ Variable, }\AttributeTok{y =}\NormalTok{ Mean,}
                  \AttributeTok{group =}\NormalTok{ Profile, }
                  \AttributeTok{shape =}\NormalTok{ Profile, }
                  \AttributeTok{color =}\NormalTok{ Profile)) }\SpecialCharTok{+}
  \FunctionTok{geom\_point}\NormalTok{(}\FunctionTok{aes}\NormalTok{(}\AttributeTok{size =}\NormalTok{ MixPro)) }\SpecialCharTok{+}
  \FunctionTok{geom\_line}\NormalTok{(}\AttributeTok{linewidth =} \FloatTok{0.5}\NormalTok{) }\SpecialCharTok{+}
  \FunctionTok{labs}\NormalTok{(}\AttributeTok{x =} \ConstantTok{NULL}\NormalTok{, }\AttributeTok{y =} \StringTok{"Latent profiles means"}\NormalTok{) }\SpecialCharTok{+}
  \FunctionTok{scale\_color\_manual}\NormalTok{(}\AttributeTok{values =} \FunctionTok{mclust.options}\NormalTok{(}\StringTok{"classPlotColors"}\NormalTok{)) }\SpecialCharTok{+}
  \FunctionTok{scale\_size}\NormalTok{(}\AttributeTok{range =} \FunctionTok{c}\NormalTok{(}\DecValTok{1}\NormalTok{, }\DecValTok{3}\NormalTok{), }\AttributeTok{guide =} \StringTok{"none"}\NormalTok{) }\SpecialCharTok{+}
  \FunctionTok{theme\_bw}\NormalTok{() }\SpecialCharTok{+}
  \FunctionTok{theme}\NormalTok{(}\AttributeTok{legend.position =} \StringTok{"top"}\NormalTok{)}
\end{Highlighting}
\end{Shaded}

\begin{figure}[H]

{\centering \includegraphics{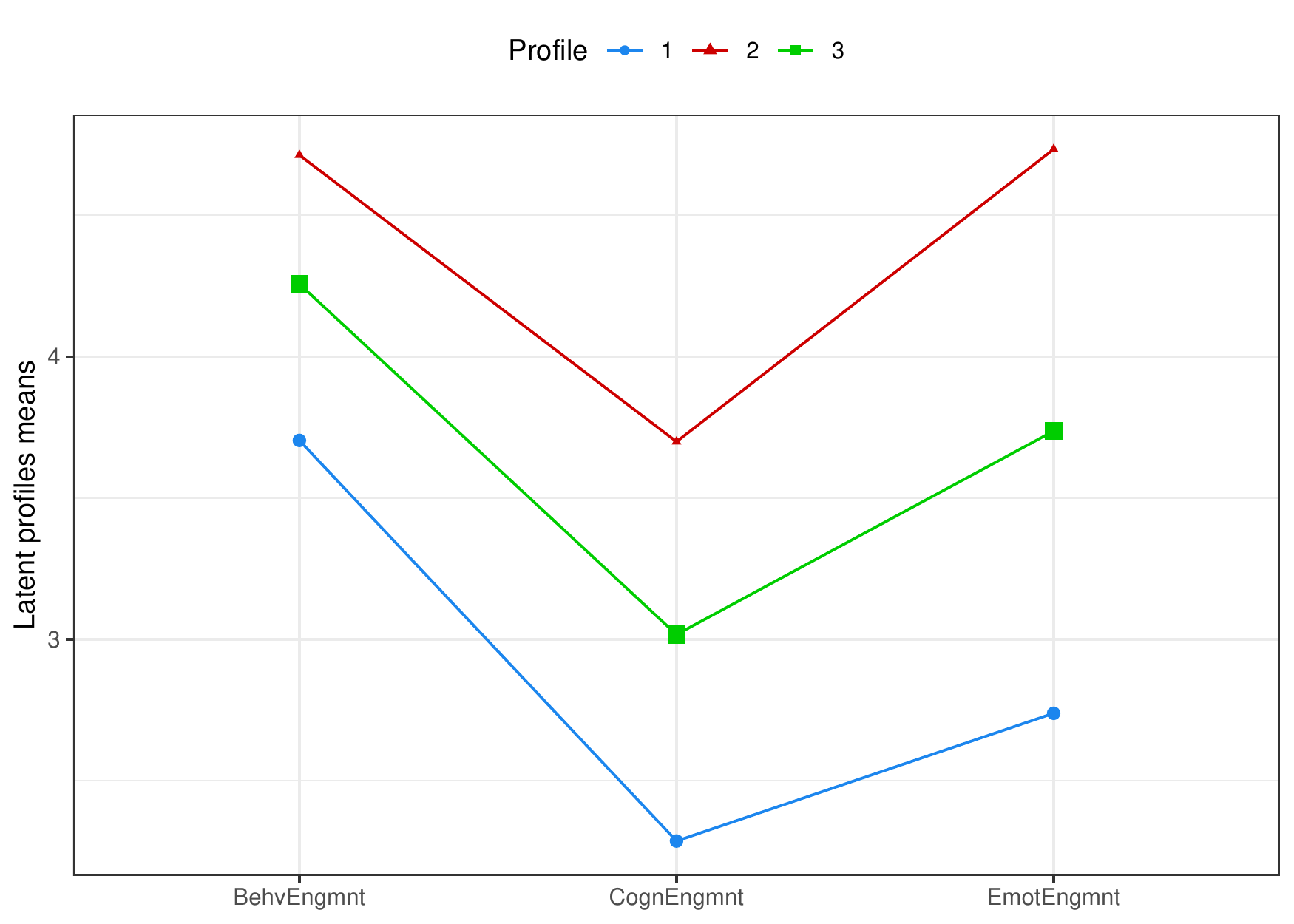}

}

\end{figure}

The smallest cluster (Profile 2) has the highest engagement scores for
all three variables. All three scores are lower for the largest cluster
(Profile 3), which are all in turn lower for Profile 1. All three
profiles exhibit the lowest mean scores for the cognitive engagement
attribute. For Profile 2, behavioural engagement and emotional
engagement scores are comparable, whereas for the other two profiles,
the mean scores for this attribute are lower than those for the
behaviour engagement attribute. Taken together, we could characterise
profiles 1, 3, and 2 as ``low'', ``medium'', and ``high'' engagement
profiles, respectively.

To provide a more comprehensive understanding of the results presented
in the previous graph, it would be beneficial to incorporate a measure
of uncertainty for the estimated means of the latent profiles. This can
be achieved by resampling using the function \texttt{MclustBootstrap()}
as described above:

\begin{Shaded}
\begin{Highlighting}[]
\NormalTok{boot }\OtherTok{\textless{}{-}} \FunctionTok{MclustBootstrap}\NormalTok{(mod, }\AttributeTok{type =} \StringTok{"bs"}\NormalTok{, }\AttributeTok{nboot =} \DecValTok{999}\NormalTok{)}
\end{Highlighting}
\end{Shaded}

The bootstrap distribution of the mixing weights can be visualized using
histograms with the code

\begin{Shaded}
\begin{Highlighting}[]
\FunctionTok{par}\NormalTok{(}\AttributeTok{mfcol =} \FunctionTok{c}\NormalTok{(}\DecValTok{1}\NormalTok{, }\DecValTok{3}\NormalTok{), }\AttributeTok{mar =} \FunctionTok{c}\NormalTok{(}\DecValTok{4}\NormalTok{, }\DecValTok{4}\NormalTok{, }\DecValTok{1}\NormalTok{, }\DecValTok{1}\NormalTok{), }\AttributeTok{mgp =} \FunctionTok{c}\NormalTok{(}\DecValTok{2}\NormalTok{, }\FloatTok{0.5}\NormalTok{, }\DecValTok{0}\NormalTok{))}
\FunctionTok{plot}\NormalTok{(boot, }\AttributeTok{what =} \StringTok{"pro"}\NormalTok{, }\AttributeTok{xlim =} \FunctionTok{c}\NormalTok{(}\DecValTok{0}\NormalTok{, }\DecValTok{1}\NormalTok{))}
\end{Highlighting}
\end{Shaded}

\begin{figure}[H]

{\centering \includegraphics{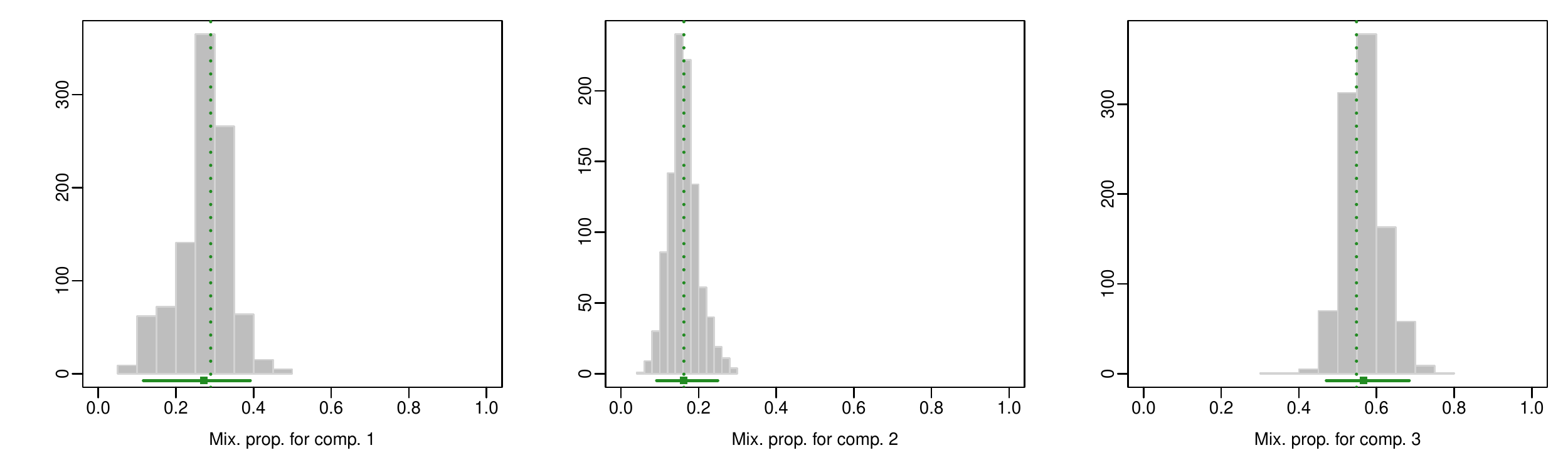}

}

\end{figure}

while the bootstrap distribution of the components means with the code

\begin{Shaded}
\begin{Highlighting}[]
\FunctionTok{par}\NormalTok{(}\AttributeTok{mfcol =} \FunctionTok{c}\NormalTok{(}\DecValTok{3}\NormalTok{, }\DecValTok{3}\NormalTok{), }\AttributeTok{mar =} \FunctionTok{c}\NormalTok{(}\DecValTok{4}\NormalTok{, }\DecValTok{4}\NormalTok{, }\DecValTok{1}\NormalTok{, }\DecValTok{1}\NormalTok{), }\AttributeTok{mgp =} \FunctionTok{c}\NormalTok{(}\DecValTok{2}\NormalTok{, }\FloatTok{0.5}\NormalTok{, }\DecValTok{0}\NormalTok{))}
\FunctionTok{plot}\NormalTok{(boot, }\AttributeTok{what =} \StringTok{"mean"}\NormalTok{, }\AttributeTok{conf.level =} \FloatTok{0.95}\NormalTok{)}
\end{Highlighting}
\end{Shaded}

\begin{figure}[H]

{\centering \includegraphics{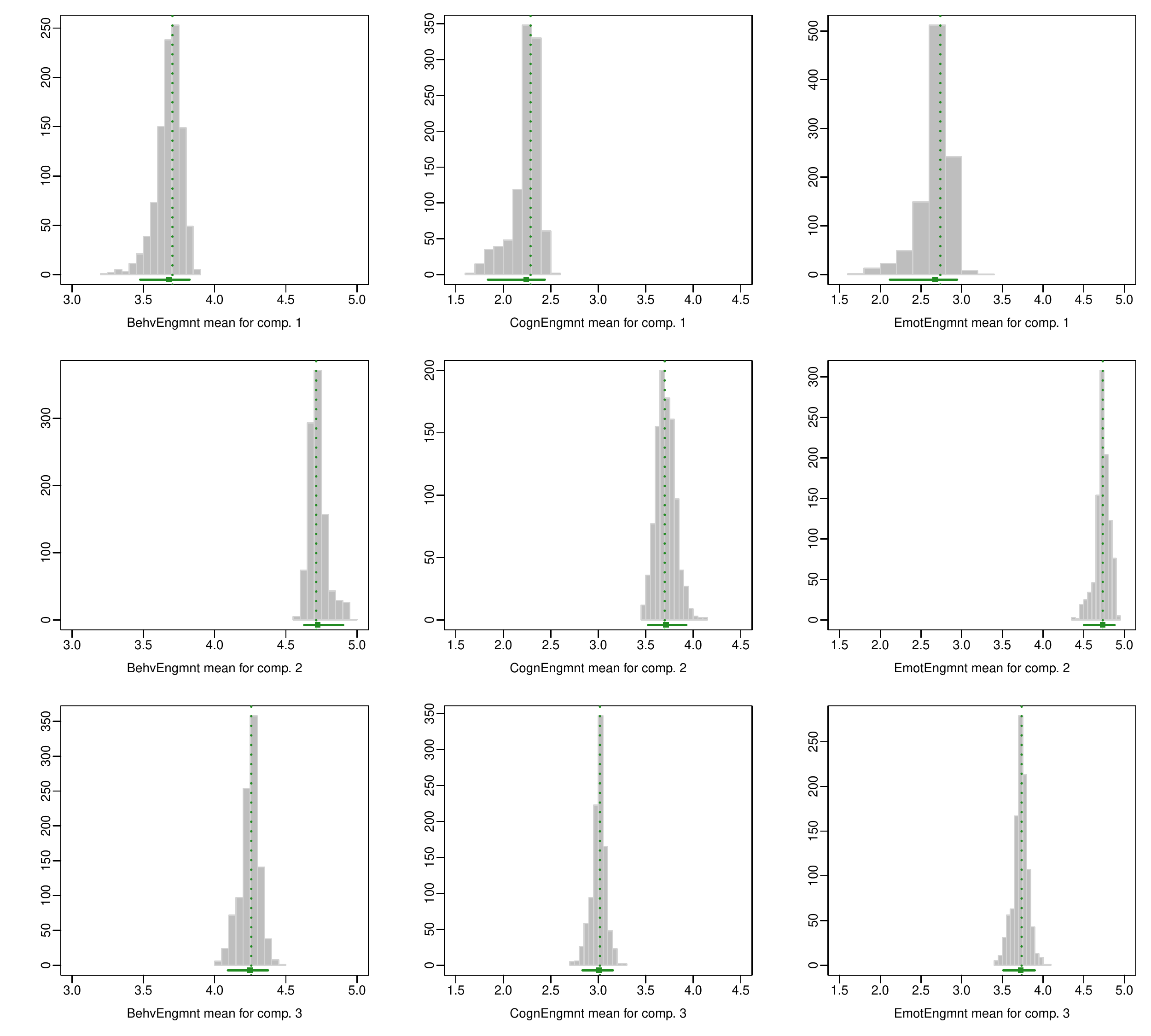}

}

\end{figure}

In all the previous graphs, the GMM estimates are shown as dashed
vertical lines, while the horizontal segments represent the percentile
confidence intervals at the 95\% confidence level.

Numerical output of the resampling-based bootstrap distributions is
obtained as:

\begin{Shaded}
\begin{Highlighting}[]
\NormalTok{sboot }\OtherTok{\textless{}{-}} \FunctionTok{summary}\NormalTok{(boot, }\AttributeTok{what =} \StringTok{"ci"}\NormalTok{)}
\NormalTok{sboot}
\DocumentationTok{\#\# {-}{-}{-}{-}{-}{-}{-}{-}{-}{-}{-}{-}{-}{-}{-}{-}{-}{-}{-}{-}{-}{-}{-}{-}{-}{-}{-}{-}{-}{-}{-}{-}{-}{-}{-}{-}{-}{-}{-}{-}{-}{-}{-}{-}{-}{-}{-}{-}{-}{-}{-}{-}{-}{-}{-}{-}{-}{-} }
\DocumentationTok{\#\# Resampling confidence intervals }
\DocumentationTok{\#\# {-}{-}{-}{-}{-}{-}{-}{-}{-}{-}{-}{-}{-}{-}{-}{-}{-}{-}{-}{-}{-}{-}{-}{-}{-}{-}{-}{-}{-}{-}{-}{-}{-}{-}{-}{-}{-}{-}{-}{-}{-}{-}{-}{-}{-}{-}{-}{-}{-}{-}{-}{-}{-}{-}{-}{-}{-}{-} }
\DocumentationTok{\#\# Model                      = VVI }
\DocumentationTok{\#\# Num. of mixture components = 3 }
\DocumentationTok{\#\# Replications               = 999 }
\DocumentationTok{\#\# Type                       = nonparametric bootstrap }
\DocumentationTok{\#\# Confidence level           = 0.95 }
\DocumentationTok{\#\# }
\DocumentationTok{\#\# Mixing probabilities:}
\DocumentationTok{\#\#               1          2         3}
\DocumentationTok{\#\# 2.5\%  0.1163335 0.09205218 0.4708275}
\DocumentationTok{\#\# 97.5\% 0.3914599 0.24915948 0.6840957}
\DocumentationTok{\#\# }
\DocumentationTok{\#\# Means:}
\DocumentationTok{\#\# [,,1]}
\DocumentationTok{\#\#       BehvEngmnt CognEngmnt EmotEngmnt}
\DocumentationTok{\#\# 2.5\%    3.477729   1.838080   2.119603}
\DocumentationTok{\#\# 97.5\%   3.823994   2.436499   2.943870}
\DocumentationTok{\#\# [,,2]}
\DocumentationTok{\#\#       BehvEngmnt CognEngmnt EmotEngmnt}
\DocumentationTok{\#\# 2.5\%    4.628794   3.525403   4.503787}
\DocumentationTok{\#\# 97.5\%   4.902464   3.927119   4.878830}
\DocumentationTok{\#\# [,,3]}
\DocumentationTok{\#\#       BehvEngmnt CognEngmnt EmotEngmnt}
\DocumentationTok{\#\# 2.5\%    4.093381   2.833736   3.512896}
\DocumentationTok{\#\# 97.5\%   4.374219   3.153141   3.898397}
\DocumentationTok{\#\# }
\DocumentationTok{\#\# Variances:}
\DocumentationTok{\#\# [,,1]}
\DocumentationTok{\#\#       BehvEngmnt CognEngmnt EmotEngmnt}
\DocumentationTok{\#\# 2.5\%   0.4003220  0.2231432  0.5528809}
\DocumentationTok{\#\# 97.5\%  0.8110118  0.4871179  0.9999560}
\DocumentationTok{\#\# [,,2]}
\DocumentationTok{\#\#       BehvEngmnt CognEngmnt EmotEngmnt}
\DocumentationTok{\#\# 2.5\%  0.01584569  0.3138691 0.01899636}
\DocumentationTok{\#\# 97.5\% 0.10470006  0.5762498 0.15014816}
\DocumentationTok{\#\# [,,3]}
\DocumentationTok{\#\#       BehvEngmnt CognEngmnt EmotEngmnt}
\DocumentationTok{\#\# 2.5\%   0.1610358  0.2874148  0.2312741}
\DocumentationTok{\#\# 97.5\%  0.2833726  0.3940830  0.4186933}
\end{Highlighting}
\end{Shaded}

The information above can then be used to plot the latent profile means
with 95\% confidence intervals shown as vertical errors bars as follows:

\begin{Shaded}
\begin{Highlighting}[]
\NormalTok{means }\OtherTok{\textless{}{-}}\NormalTok{ means }\SpecialCharTok{|\textgreater{}} 
  \FunctionTok{add\_column}\NormalTok{(}\AttributeTok{lower =} \FunctionTok{as.vector}\NormalTok{(sboot}\SpecialCharTok{$}\NormalTok{mean[}\DecValTok{1}\NormalTok{,,]),}
             \AttributeTok{upper =} \FunctionTok{as.vector}\NormalTok{(sboot}\SpecialCharTok{$}\NormalTok{mean[}\DecValTok{2}\NormalTok{,,]))}
\NormalTok{means}
\DocumentationTok{\#\# \# A tibble: 9 x 6}
\DocumentationTok{\#\#   Profile Variable    Mean MixPro lower upper}
\DocumentationTok{\#\#   \textless{}fct\textgreater{}   \textless{}fct\textgreater{}      \textless{}dbl\textgreater{}  \textless{}dbl\textgreater{} \textless{}dbl\textgreater{} \textless{}dbl\textgreater{}}
\DocumentationTok{\#\# 1 1       BehvEngmnt  3.70  0.290  3.48  3.82}
\DocumentationTok{\#\# 2 1       CognEngmnt  2.29  0.290  1.84  2.44}
\DocumentationTok{\#\# 3 1       EmotEngmnt  2.74  0.290  2.12  2.94}
\DocumentationTok{\#\# 4 2       BehvEngmnt  4.71  0.162  4.63  4.90}
\DocumentationTok{\#\# 5 2       CognEngmnt  3.70  0.162  3.53  3.93}
\DocumentationTok{\#\# 6 2       EmotEngmnt  4.73  0.162  4.50  4.88}
\DocumentationTok{\#\# 7 3       BehvEngmnt  4.26  0.548  4.09  4.37}
\DocumentationTok{\#\# 8 3       CognEngmnt  3.02  0.548  2.83  3.15}
\DocumentationTok{\#\# 9 3       EmotEngmnt  3.74  0.548  3.51  3.90}

\FunctionTok{ggplot}\NormalTok{(means, }\FunctionTok{aes}\NormalTok{(}\AttributeTok{x =}\NormalTok{ Variable, }\AttributeTok{y =}\NormalTok{ Mean, }\AttributeTok{group =}\NormalTok{ Profile, }
                  \AttributeTok{shape =}\NormalTok{ Profile, }\AttributeTok{color =}\NormalTok{ Profile)) }\SpecialCharTok{+}
  \FunctionTok{geom\_point}\NormalTok{(}\FunctionTok{aes}\NormalTok{(}\AttributeTok{size =}\NormalTok{ MixPro)) }\SpecialCharTok{+}
  \FunctionTok{geom\_line}\NormalTok{(}\AttributeTok{linewidth =} \FloatTok{0.5}\NormalTok{) }\SpecialCharTok{+}
  \FunctionTok{geom\_errorbar}\NormalTok{(}\FunctionTok{aes}\NormalTok{(}\AttributeTok{ymin =}\NormalTok{ lower, }\AttributeTok{ymax =}\NormalTok{ upper), }
                \AttributeTok{linewidth =} \FloatTok{0.5}\NormalTok{, }\AttributeTok{width =} \FloatTok{0.1}\NormalTok{) }\SpecialCharTok{+}
  \FunctionTok{labs}\NormalTok{(}\AttributeTok{x =} \ConstantTok{NULL}\NormalTok{, }\AttributeTok{y =} \StringTok{"Latent profiles means"}\NormalTok{) }\SpecialCharTok{+}
  \FunctionTok{scale\_color\_manual}\NormalTok{(}\AttributeTok{values =} \FunctionTok{mclust.options}\NormalTok{(}\StringTok{"classPlotColors"}\NormalTok{)) }\SpecialCharTok{+}
  \FunctionTok{scale\_size}\NormalTok{(}\AttributeTok{range =} \FunctionTok{c}\NormalTok{(}\DecValTok{1}\NormalTok{, }\DecValTok{3}\NormalTok{), }\AttributeTok{guide =} \StringTok{"none"}\NormalTok{) }\SpecialCharTok{+}
  \FunctionTok{theme\_bw}\NormalTok{() }\SpecialCharTok{+}
  \FunctionTok{theme}\NormalTok{(}\AttributeTok{legend.position =} \StringTok{"top"}\NormalTok{)}
\end{Highlighting}
\end{Shaded}

\begin{figure}[H]

{\centering \includegraphics{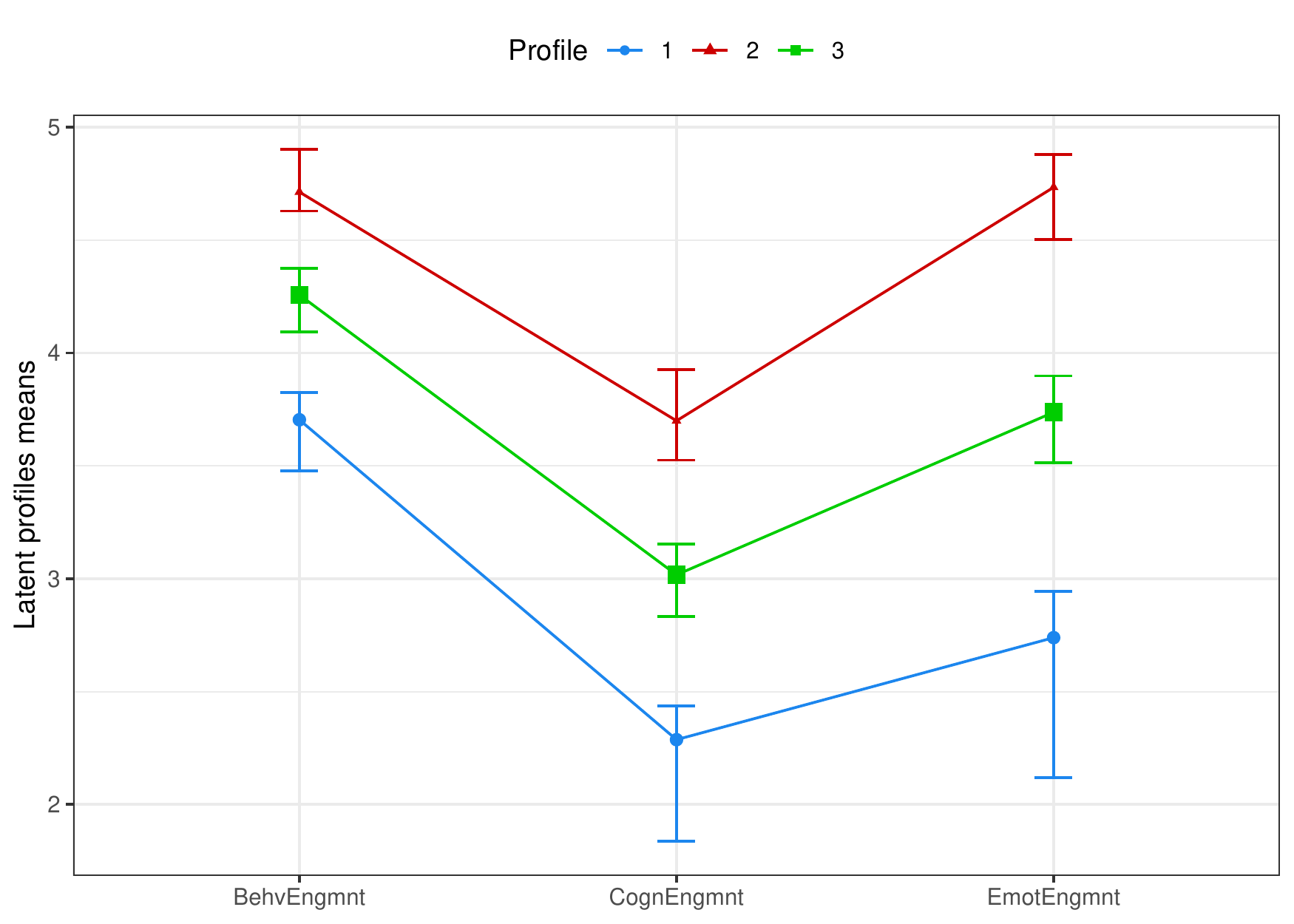}

}

\end{figure}

The error bars for the cognitive and emotional engagement attributes are
noticeably wider for the ``low'' engagement profile.

Finally, the entropy of the estimated classification, average entropy of
each latent component, and average posterior probabilities are obtained
via:

\begin{Shaded}
\begin{Highlighting}[]
\NormalTok{probs }\OtherTok{\textless{}{-}}\NormalTok{ mod}\SpecialCharTok{$}\NormalTok{z                    }\CommentTok{\# posterior conditional probs}
\NormalTok{probs\_map }\OtherTok{\textless{}{-}} \FunctionTok{apply}\NormalTok{(probs, }\DecValTok{1}\NormalTok{, max) }\CommentTok{\# maximum a posteriori probs}
\NormalTok{class }\OtherTok{\textless{}{-}}\NormalTok{ mod}\SpecialCharTok{$}\NormalTok{classification       }\CommentTok{\# latent classes for each obs}
\NormalTok{n }\OtherTok{\textless{}{-}}\NormalTok{ mod}\SpecialCharTok{$}\NormalTok{n                        }\CommentTok{\# number of obs}
\NormalTok{K }\OtherTok{\textless{}{-}}\NormalTok{ mod}\SpecialCharTok{$}\NormalTok{G                        }\CommentTok{\# number of latent classes}

\CommentTok{\# Entropy}
\NormalTok{E }\OtherTok{\textless{}{-}} \DecValTok{1} \SpecialCharTok{+} \FunctionTok{sum}\NormalTok{(probs }\SpecialCharTok{*} \FunctionTok{log}\NormalTok{(probs))}\SpecialCharTok{/}\NormalTok{(n }\SpecialCharTok{*} \FunctionTok{log}\NormalTok{(K))}
\NormalTok{E}
\DocumentationTok{\#\# [1] 0.6890602}

\CommentTok{\# Case{-}specific entropy contributions}
\NormalTok{Ei }\OtherTok{\textless{}{-}} \DecValTok{1} \SpecialCharTok{+} \FunctionTok{rowSums}\NormalTok{(probs }\SpecialCharTok{*} \FunctionTok{log}\NormalTok{(probs))}\SpecialCharTok{/}\FunctionTok{log}\NormalTok{(K)}
\FunctionTok{sum}\NormalTok{(Ei)}\SpecialCharTok{/}\NormalTok{n}
\DocumentationTok{\#\# [1] 0.6890602}

\NormalTok{df\_entropy  }\OtherTok{\textless{}{-}} \FunctionTok{data.frame}\NormalTok{(}\AttributeTok{class =} \FunctionTok{as.factor}\NormalTok{(class), }\AttributeTok{entropy =}\NormalTok{ Ei)}

\NormalTok{df\_entropy }\SpecialCharTok{|\textgreater{}}
  \FunctionTok{group\_by}\NormalTok{(class) }\SpecialCharTok{|\textgreater{}}
  \FunctionTok{summarise}\NormalTok{(}\AttributeTok{count =} \FunctionTok{n}\NormalTok{(),}
            \AttributeTok{mean =} \FunctionTok{mean}\NormalTok{(entropy),}
            \AttributeTok{sd =} \FunctionTok{sd}\NormalTok{(entropy),}
            \AttributeTok{min =} \FunctionTok{min}\NormalTok{(entropy),}
            \AttributeTok{max =} \FunctionTok{max}\NormalTok{(entropy))}
\DocumentationTok{\#\# \# A tibble: 3 x 6}
\DocumentationTok{\#\#   class count  mean    sd   min   max}
\DocumentationTok{\#\#   \textless{}fct\textgreater{} \textless{}int\textgreater{} \textless{}dbl\textgreater{} \textless{}dbl\textgreater{} \textless{}dbl\textgreater{} \textless{}dbl\textgreater{}}
\DocumentationTok{\#\# 1 1       184 0.740 0.239 0.369 1.00 }
\DocumentationTok{\#\# 2 2       119 0.690 0.225 0.187 0.993}
\DocumentationTok{\#\# 3 3       414 0.666 0.197 0.172 0.974}

\FunctionTok{ggplot}\NormalTok{(df\_entropy, }\FunctionTok{aes}\NormalTok{(}\AttributeTok{y =}\NormalTok{ class, }\AttributeTok{x =}\NormalTok{ entropy,  }\AttributeTok{fill =}\NormalTok{ class)) }\SpecialCharTok{+}
  \FunctionTok{geom\_density\_ridges}\NormalTok{(}\AttributeTok{stat =} \StringTok{"binline"}\NormalTok{, }\AttributeTok{bins =} \DecValTok{21}\NormalTok{,}
                      \AttributeTok{scale =} \FloatTok{0.9}\NormalTok{, }\AttributeTok{alpha =} \FloatTok{0.5}\NormalTok{) }\SpecialCharTok{+}
  \FunctionTok{scale\_x\_continuous}\NormalTok{(}\AttributeTok{breaks =} \FunctionTok{seq}\NormalTok{(}\DecValTok{0}\NormalTok{, }\DecValTok{1}\NormalTok{ ,}\AttributeTok{by=}\FloatTok{0.1}\NormalTok{), }
                     \AttributeTok{limits =} \FunctionTok{c}\NormalTok{(}\DecValTok{0}\NormalTok{, }\FloatTok{1.05}\NormalTok{)) }\SpecialCharTok{+}
  \FunctionTok{scale\_fill\_manual}\NormalTok{(}\AttributeTok{values =} \FunctionTok{mclust.options}\NormalTok{(}\StringTok{"classPlotColors"}\NormalTok{)) }\SpecialCharTok{+}
  \FunctionTok{geom\_vline}\NormalTok{(}\AttributeTok{xintercept =}\NormalTok{ E, }\AttributeTok{lty =} \DecValTok{2}\NormalTok{) }\SpecialCharTok{+}
  \FunctionTok{labs}\NormalTok{(}\AttributeTok{x =} \StringTok{"Case{-}specific entropy contribution"}\NormalTok{, }
       \AttributeTok{y =} \StringTok{"Latent class"}\NormalTok{) }\SpecialCharTok{+}
  \FunctionTok{theme\_ridges}\NormalTok{(}\AttributeTok{center\_axis\_labels =} \ConstantTok{TRUE}\NormalTok{) }\SpecialCharTok{+}
  \FunctionTok{theme}\NormalTok{(}\AttributeTok{legend.position =} \StringTok{"none"}\NormalTok{,}
        \AttributeTok{panel.spacing =} \FunctionTok{unit}\NormalTok{(}\DecValTok{1}\NormalTok{, }\StringTok{"lines"}\NormalTok{),}
        \AttributeTok{strip.text.x =} \FunctionTok{element\_text}\NormalTok{(}\AttributeTok{size =} \DecValTok{8}\NormalTok{)) }
\end{Highlighting}
\end{Shaded}

\begin{figure}[H]

{\centering \includegraphics{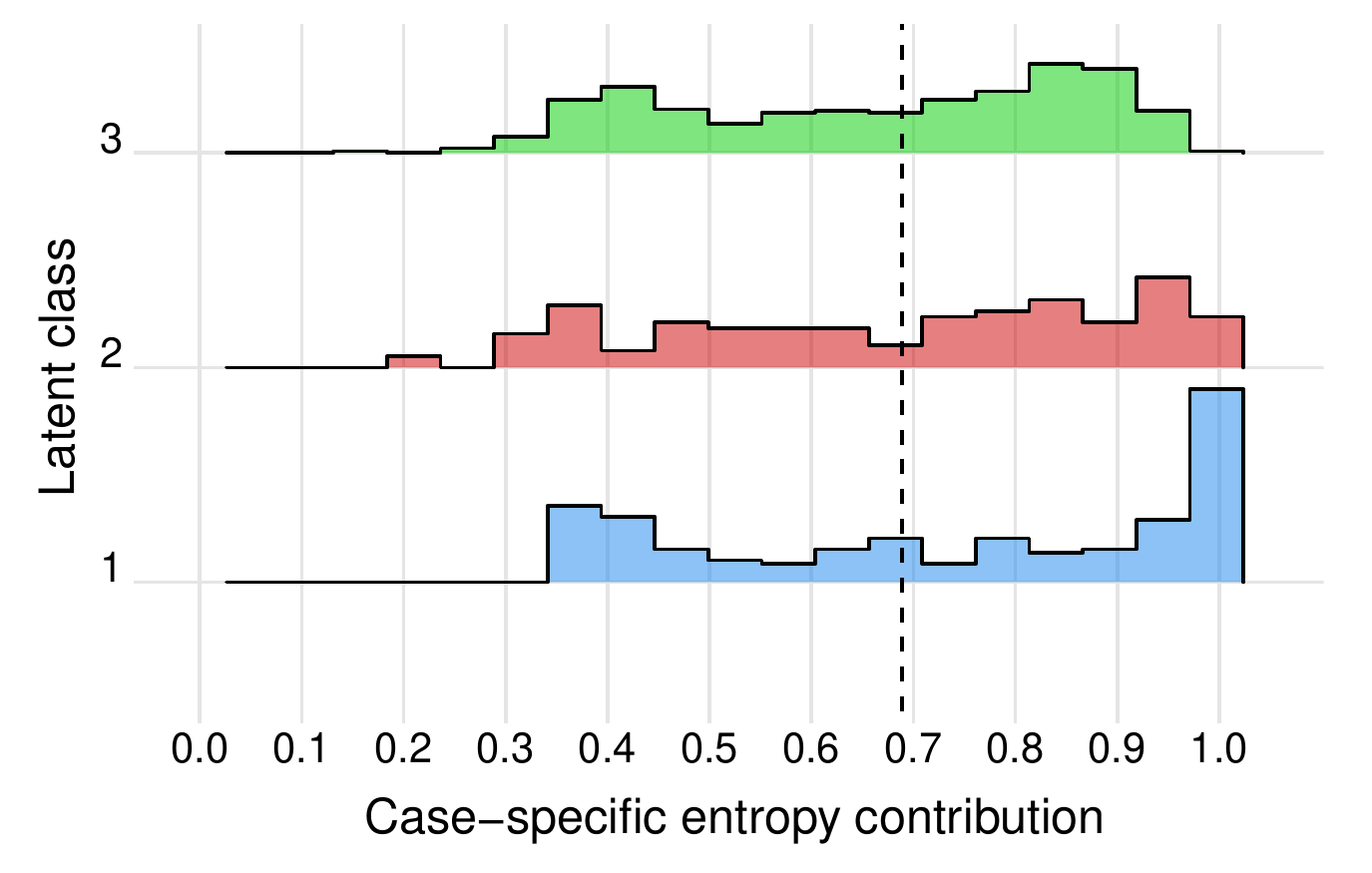}

}

\end{figure}

\begin{Shaded}
\begin{Highlighting}[]
\CommentTok{\# Average posterior probabilities by latent class:}
\NormalTok{df\_AvePP }\OtherTok{\textless{}{-}} \FunctionTok{data.frame}\NormalTok{(}\AttributeTok{class =} \FunctionTok{as.factor}\NormalTok{(class), }\AttributeTok{pp =}\NormalTok{ probs\_map)}

\NormalTok{df\_AvePP }\SpecialCharTok{|\textgreater{}}
  \FunctionTok{group\_by}\NormalTok{(class) }\SpecialCharTok{|\textgreater{}}
  \FunctionTok{summarise}\NormalTok{(}\AttributeTok{count =} \FunctionTok{n}\NormalTok{(),}
            \AttributeTok{mean =} \FunctionTok{mean}\NormalTok{(pp),}
            \AttributeTok{sd =} \FunctionTok{sd}\NormalTok{(pp),}
            \AttributeTok{min =} \FunctionTok{min}\NormalTok{(pp),}
            \AttributeTok{max =} \FunctionTok{max}\NormalTok{(pp))}
\DocumentationTok{\#\# \# A tibble: 3 x 6}
\DocumentationTok{\#\#   class count  mean    sd   min   max}
\DocumentationTok{\#\#   \textless{}fct\textgreater{} \textless{}int\textgreater{} \textless{}dbl\textgreater{} \textless{}dbl\textgreater{} \textless{}dbl\textgreater{} \textless{}dbl\textgreater{}}
\DocumentationTok{\#\# 1 1       184 0.864 0.160 0.513 1.00 }
\DocumentationTok{\#\# 2 2       119 0.858 0.146 0.468 0.999}
\DocumentationTok{\#\# 3 3       414 0.850 0.135 0.502 0.996}

\FunctionTok{ggplot}\NormalTok{(df\_AvePP, }\FunctionTok{aes}\NormalTok{(}\AttributeTok{y =}\NormalTok{ class, }\AttributeTok{x =}\NormalTok{ pp,  }\AttributeTok{fill =}\NormalTok{ class)) }\SpecialCharTok{+}
  \FunctionTok{geom\_density\_ridges}\NormalTok{(}\AttributeTok{stat =} \StringTok{"binline"}\NormalTok{, }\AttributeTok{bins =} \DecValTok{21}\NormalTok{,}
                      \AttributeTok{scale =} \FloatTok{0.9}\NormalTok{, }\AttributeTok{alpha =} \FloatTok{0.5}\NormalTok{) }\SpecialCharTok{+}
  \FunctionTok{scale\_x\_continuous}\NormalTok{(}\AttributeTok{breaks =} \FunctionTok{seq}\NormalTok{(}\DecValTok{0}\NormalTok{, }\DecValTok{1}\NormalTok{, }\AttributeTok{by=}\FloatTok{0.1}\NormalTok{), }
                     \AttributeTok{limits =} \FunctionTok{c}\NormalTok{(}\DecValTok{0}\NormalTok{, }\FloatTok{1.05}\NormalTok{)) }\SpecialCharTok{+}
  \FunctionTok{scale\_fill\_manual}\NormalTok{(}\AttributeTok{values =} \FunctionTok{mclust.options}\NormalTok{(}\StringTok{"classPlotColors"}\NormalTok{)) }\SpecialCharTok{+}
  \FunctionTok{labs}\NormalTok{(}\AttributeTok{x =} \StringTok{"MAP probabilities"}\NormalTok{, }\AttributeTok{y =} \StringTok{"Latent class"}\NormalTok{) }\SpecialCharTok{+}
  \FunctionTok{theme\_ridges}\NormalTok{(}\AttributeTok{center\_axis\_labels =} \ConstantTok{TRUE}\NormalTok{) }\SpecialCharTok{+}
  \FunctionTok{theme}\NormalTok{(}\AttributeTok{legend.position =} \StringTok{"none"}\NormalTok{,}
        \AttributeTok{panel.spacing =} \FunctionTok{unit}\NormalTok{(}\DecValTok{1}\NormalTok{, }\StringTok{"lines"}\NormalTok{),}
        \AttributeTok{strip.text.x =} \FunctionTok{element\_text}\NormalTok{(}\AttributeTok{size =} \DecValTok{8}\NormalTok{)) }
\end{Highlighting}
\end{Shaded}

\begin{figure}[H]

{\centering \includegraphics{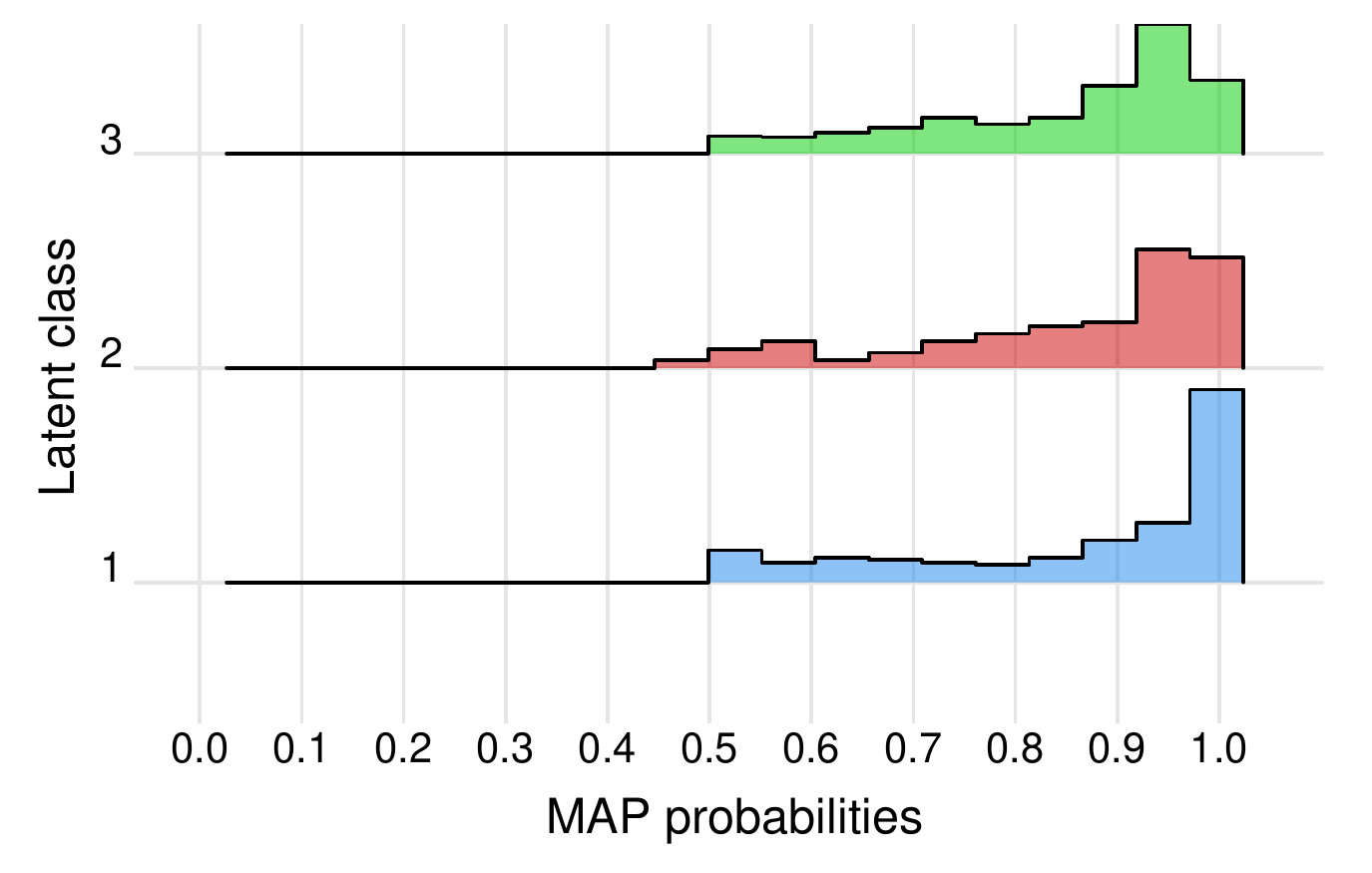}

}

\end{figure}

We note that all entropy and AvePP quantities appear satisfactory from
the point of view of indicating reasonably well-separated clusters.

\hypertarget{discussion}{%
\section{Discussion}\label{discussion}}

Using a person-centered method (finite Gaussian mixture model), the
present analysis uncovered the heterogeneity within the SEM engagement
data by identifying three latent or unobserved clusters: low, medium,
and high engagement clusters. Uncovering the latent structure could help
understand individual differences among students, identify the complex
multidimensional variability of a construct ---engagement in our case---
and possibly help personalize teaching and learning. Several studies
have revealed similar patterns of engagement which ---similar to the
current analysis--- comprise three levels that can be roughly summarized
as high, moderate, and low (Saqr and López-Pernas 2021; Archambault and
Dupéré 2016; Zhen et al. 2019). The heterogeneity of engagement has been
demonstrated in longitudinal studies, in both face-to-face settings as
well as online engagement (Saqr and López-Pernas 2021). Furthermore, the
association between engagement and performance has been demonstrated to
vary by achievement level, time of the year, as well as engagement
state; that is, high achievers may at some point in their program
descend to lower engagement states and still continue to have higher
achievement (Saqr et al. 2023). Such patterns, variability, and
individual differences are not limited to engagement, but has been
reported for almost every major disposition in education psychology
(Hickendorff et al. 2018).

On a general level, heterogeneity has been a hot topic in recent
educational literature. Several calls have been voiced to adopt methods
that capture different patterns or subgroups within students' behavior
or functioning. Assuming that there is ``an average'' pattern that
represents the entirety of student populations requires the measured
construct to have the same causal mechanism, same development pattern,
and affect students in exactly the same way. The average assumption is
of course impossible and has been proven inaccurate across a vast number
of studies (e.g., Hickendorff et al. (2018) and Törmänen et al. (2022)).
Since heterogeneity is prevalent in psychological, behavioral, and
physiological human data, person-centered methods will remain a very
important tool for researchers (Bryan, Tipton, and Yeager 2021).

Person-centered methods can be grouped into traditional, algorithmic
clustering methods on one hand and the model-based clustering paradigm
on the other. The analysis of the SEM data centered here on the
model-based approach, specifically the finite Gaussian mixture model
framework. The \textbf{mclust} package enabled such models to be fitted
quickly and easily and this framework exhibits many advantages over
traditional clustering algorithms which rely on distance-based
heuristics. Firstly, the likelihood-based underpinnings enable the
selection of the optimal model using principled statistical model
selection criteria. In particular, it is noteworthy in the present
analysis that the model selection procedure was not limited to
three-cluster solutions: mixtures with fewer or greater than three
clusters were evaluated and the three-cluster solution ---supported by
previous studies in education research--- was identified as optimal
according to the BIC. Secondly, the parsimonious modelling of the
covariance structures provides the flexibility to model clusters with
different geometric characteristics. In particular, the clusters in the
present analysis, whereby each group is described by a single Gaussian
component with varying volume and shape, but the same orientation
aligned with the coordinate axes are more flexible than the spherical,
Euclidean distance-based clusters obtainable under the \(k\)-means
algorithm. Thirdly, the models relax the assumption that each
observation is associated with exactly one cluster and yields
informative cluster-membership probabilities for each observation, which
can be used to compute useful diagnostics such as entropies and average
posterior probabilities which are unavailable under so-called ``hard''
clustering frameworks. Finally, the \textbf{mclust} package facilitates
simple summaries and visualisations of the resulting clusters and
cluster-specific parameter estimates.

That being said, there are a number of methodological limitations of the
GMM framework to be aware of in other settings. Firstly, and most
obviously, such models are inappropriate for clustering categorical or
mixed-type variables. \textbf{For clustering longitudinal categorical
sequences, such as those in Chapter XXX}, model-based approaches are
provided by the mixtures of exponential-distance models framework of
Murphy et al. (2021) (and the associated \texttt{MEDseq}) R package and
the mixtures of hidden Markov models framework of Helske and Helske
(2019) (and the associated \texttt{seqHMM}\} package). Regarding
mixed-type variables, McParland and Gormley (2016) provide a model-based
framework (and the associated \texttt{clustMD} package.

Secondly, the one-to-one correspondence typically assumed between
component distributions and clusters is is not always the case (Hennig
2010). This is only true if the underlying true component densities are
Gaussian. When the assumption of component-wise normality is not
satisfied, the performance of such models will deteriorate as more
components are required to fit the data well. However, even for
continuous data, GMMs tend to overestimate the number of clusters when
the assumption of normality is violated. Two strategies for dealing with
this are provided by the \textbf{mclust} package, one based on combining
Gaussian mixture components according to an entropy criterion, and one
based on a adding a so-called ``noise component'' ---represented by a
uniform distribution--- to the mixture. The noise component captures
outliers with do not fit the prevailing patterns of Gaussian clusters,
which would otherwise be assigned to (possibly many) small clusters and
minimises their deleterious effect on parameter estimation for the
other, more defined clusters. Further details of combining components
and adding a noise component can be found in Scrucca et al. (2023, chap.
7). Alternatively, mixture models which depart from normality have been
an active area of research in model-based clustering in recent years.
Such approaches ---some of which are available in the R package
\textbf{mixture} (Pocuca, Browne, and McNicholas 2022)--- replace the
underlying Gaussian component distributions with e.g., generalised
hyperbolic distributions, the multivariate \(t\) distribution, and the
multivarate skew-\(t\) distribution.

A third main limitation of GMMs is their ineffectiveness in
high-dimensional settings, when the data dimension \(d\) is comparable
to or even greater than \(n\). Among the 14 parsimonious
parameterisations available in \textbf{mclust}, only models with
diagonal covariance structures are tractable when \(n \le p\).
Incorporating factor-analytic covariance decompositions in so-called
finite Gaussian mixtures of factor analysers have been proposed for
addressing this issue (Ghahramani and Hinton 1996; McLachlan, Peel, and
Bean 2003). Imposing constraints on the parameters of such
factor-analytic structures in the component covariance matrices in the
spirit of \textbf{mclust} leads to another family of parsimonious
Gaussian mixture models McNicholas and Murphy (2008), which are
implemented in the R package \textbf{pgmm}. Model selection becomes
increasingly difficult with such models, given the need to choose both
the optimal number of mixture components and the optimal number of
latent factors (as well as the covariance parameterisation, in the case
of \textbf{pgmm}). Infinite mixtures of infinite factor analysers
---implemented in the R package \textbf{IMIFA}--- are a recent, Bayesian
extension which enable automatic inference of the number of components
and the numbers of cluster-specific latent factors (Murphy, Viroli, and
Gormley 2020).

Another recent extension, building directly on the 14 models from
\textbf{mclust}, is the MoEClust model family of Murphy and Murphy
(2020) and the associated \textbf{MoEClust} R package, which closely
mimics its syntax. MoEClust effectively embeds Gaussian parsimonious
clustering models in the mixtures of experts framework, enabling
additional sources of heterogeneity in the form of covariates to be
incorporated directly in the clustering model, to guide the construction
of the clusters. Either, neither, or both the mixing proportions and/or
component mean parameters can be modelled as functions of these
covariates. The former is perhaps particularly appealing, given its
analogous equivalence to latent profile \emph{regression} (Dayton and
Macready 1988). Hypothetically, assuming information on the gender and
age of the students in the present analysis was available, such
covariates would influence the probabilities of cluster membership under
such a model, while the correspondence thereafter between the parameters
of the component distributions and the clusters would have the same
interpretation as per standard LPA models.

\hypertarget{references}{%
\section{References}\label{references}}

\hypertarget{refs}{}
\begin{CSLReferences}{1}{0}
\leavevmode\vadjust pre{\hypertarget{ref-Archambault2016}{}}%
Archambault, Isabelle, and Véronique Dupéré. 2016. {``Joint Trajectories
of Behavioral, Affective, and Cognitive Engagement in Elementary
School.''} \emph{The Journal of Educational Research} 110 (2): 188--98.
\url{https://doi.org/10.1080/00220671.2015.1060931}.

\leavevmode\vadjust pre{\hypertarget{ref-Banfield:Raftery:1993}{}}%
Banfield, J., and Adrian E. Raftery. 1993. {``Model-Based {G}aussian and
Non-{G}aussian Clustering.''} \emph{Biometrics} 49 (3): 803--21.
\url{https://doi.org/10.2307/2532201}.

\leavevmode\vadjust pre{\hypertarget{ref-Bartholomew:Knott:Moustaki:2011}{}}%
Bartholomew, David J, Martin Knott, and Irini Moustaki. 2011.
\emph{Latent Variable Models and Factor Analysis: {A} Unified Approach}.
3rd ed. Vol. 904. Wiley Series in Probability and Statistics.
Chichester, UK: John Wiley \& Sons.

\leavevmode\vadjust pre{\hypertarget{ref-Basford:Greenway:McLachlan:Peel:1997}{}}%
Basford, K E, D R Greenway, G J McLachlan, and D Peel. 1997. {``Standard
Errors of Fitted Component Means of Normal Mixtures.''}
\emph{Computational Statistics} 12 (1): 1--18.

\leavevmode\vadjust pre{\hypertarget{ref-Biernacki:Celeux:Govaert:2000}{}}%
Biernacki, Christophe, Gilles Celeux, and Gérard Govaert. 2000.
{``Assessing a Mixture Model for Clustering with the Integrated
Completed Likelihood.''} \emph{{IEEE} Transactions on Pattern Analysis
and Machine Intelligence} 22 (7): 719--25.

\leavevmode\vadjust pre{\hypertarget{ref-Blei:Ng:Jordan:2003}{}}%
Blei, David M, Andrew Y Ng, and Michael I Jordan. 2003. {``Latent
{D}irichlet Allocation.''} \emph{Journal of Machine Learning Research}
3: 993--1022.

\leavevmode\vadjust pre{\hypertarget{ref-Bryan2021}{}}%
Bryan, Christopher J., Elizabeth Tipton, and David S. Yeager. 2021.
{``Behavioural Science Is Unlikely to Change the World Without a
Heterogeneity Revolution.''} \emph{Nature Human Behaviour} 5 (8):
980--89. \url{https://doi.org/10.1038/s41562-021-01143-3}.

\leavevmode\vadjust pre{\hypertarget{ref-Celeux:Govaert:1995}{}}%
Celeux, Gilles, and Gérard Govaert. 1995. {``{G}aussian Parsimonious
Clustering Models.''} \emph{Pattern Recognition} 28 (5): 781--93.
\url{https://doi.org/10.1016/0031-3203(94)00125-6}.

\leavevmode\vadjust pre{\hypertarget{ref-Celeux:Soromenho:1996}{}}%
Celeux, Gilles, and Gilda Soromenho. 1996. {``An Entropy Criterion for
Assessing the Number of Clusters in a Mixture Model.''} \emph{Journal of
Classification} 13 (2): 195--212.

\leavevmode\vadjust pre{\hypertarget{ref-Cheng2023-fc}{}}%
Cheng, Shonn, Jui-Chieh Huang, and Waneta Hebert. 2023. {``Profiles of
Vocational College Students' Achievement Emotions in Online Learning
Environments: {A}ntecedents and Outcomes.''} \emph{Computers in Human
Behaviour} 138: 107452. \url{https://doi.org/10.1016/j.chb.2022.107452}.

\leavevmode\vadjust pre{\hypertarget{ref-Cover:Thomas:2006}{}}%
Cover, Thomas M, and Joy A Thomas. 2006. \emph{Elements of Information
Theory}. 2nd ed. Vol. 20. Wiley Series in Telecommunications and Signal
Processing. New York, NY, USA: John Wiley \& Sons.

\leavevmode\vadjust pre{\hypertarget{ref-Dayton1988}{}}%
Dayton, C. M., and G. B. Macready. 1988. {``{C}oncomitant-Variable
Latent-Class Models.''} \emph{Journal of the American Statistical
Association} 83 (401): 173--78.

\leavevmode\vadjust pre{\hypertarget{ref-Dempster:Laird:Rubin:1977}{}}%
Dempster, A. P., N. M. Laird, and D. B. Rubin. 1977. {``Maximum
Likelihood from Incomplete Data via the {EM} Algorithm (with
Discussion).''} \emph{Journal of the Royal Statistical Society: Series B
(Statistical Methodology)} 39 (1): 1--38.
\url{https://doi.org/10.1111/j.2517-6161.1977.tb01600.x}.

\leavevmode\vadjust pre{\hypertarget{ref-Efron:1979}{}}%
Efron, Bradley. 1979. {``Bootstrap Methods: {A}nother Look at the
Jackknife.''} \emph{The Annals of Statistics} 7 (1): 1--26.

\leavevmode\vadjust pre{\hypertarget{ref-Everitt2011}{}}%
Everitt, B. S., S. Landau, M. Leese, and D. Stahl. 2011. \emph{{Cluster
Analysis}}. Fifth. Vol. 848. Wiley Series in Probability and Statistics.
New York, NY, USA: John Wiley \& Sons.

\leavevmode\vadjust pre{\hypertarget{ref-Fraley:Raftery:2002}{}}%
Fraley, Chris, and Adrian E. Raftery. 2002. {``Model-Based Clustering,
Discriminant Analysis, and Density Estimation.''} \emph{Journal of the
American Statistical Association} 97 (458): 611--31.
\url{https://doi.org/10.1198/016214502760047131}.

\leavevmode\vadjust pre{\hypertarget{ref-Fraley:Raftery:2007a}{}}%
---------. 2007. {``Bayesian Regularization for Normal Mixture
Estimation and Model-Based Clustering.''} \emph{Journal of
Classification} 24 (2): 155--81.

\leavevmode\vadjust pre{\hypertarget{ref-Rpkg:mclust}{}}%
Fraley, Chris, Adrian E. Raftery, and Luca Scrucca. 2023.
\emph{{mclust}: {G}aussian Mixture Modelling for Model-Based Clustering,
Classification, and Density Estimation}.
\url{https://CRAN.R-project.org/package=mclust}.

\leavevmode\vadjust pre{\hypertarget{ref-Ghahramani1996}{}}%
Ghahramani, Z., and G. E. Hinton. 1996. {``The {EM} Algorithm for
Mixtures of Factor Analyzers.''} Department of Computer Science,
University of Toronto.

\leavevmode\vadjust pre{\hypertarget{ref-seqHMM2019}{}}%
Helske, S., and J. Helske. 2019. {``{M}ixture Hidden {M}arkov Models for
Sequence Data: {T}he {seqHMM} Package in {R}.''} \emph{Journal of
Statistical Software} 88 (3): 1--32.

\leavevmode\vadjust pre{\hypertarget{ref-Hennig2010}{}}%
Hennig, C. 2010. {``Methods for Merging {G}aussian Mixture
Components.''} \emph{Advances in Data Analysis and Classification} 4
(1): 3--34.

\leavevmode\vadjust pre{\hypertarget{ref-Hennig2015}{}}%
---------. 2015. {``What Are the True Clusters?''} \emph{Pattern
Recognition Letters} 64: 53--62.

\leavevmode\vadjust pre{\hypertarget{ref-Hickendorff2018-kt}{}}%
Hickendorff, Marian, Peter A Edelsbrunner, Jake McMullen, Michael
Schneider, and Kelly Trezise. 2018. {``Informative Tools for
Characterizing Individual Differences in Learning: {L}atent Class,
Latent Profile, and Latent Transition Analysis.''} \emph{Learning and
Individual Differences} 66: 4--15.
\url{https://doi.org/10.1016/j.lindif.2017.11.001}.

\leavevmode\vadjust pre{\hypertarget{ref-Hoi2023-sz}{}}%
Hoi, Vo Ngoc. 2023. {``Transitioning from School to University: {A}
Person-Oriented Approach to Understanding First-Year Students' Classroom
Engagement in Higher Education.''} \emph{Educational Review}, 1--21.
\url{https://doi.org/10.1080/00131911.2022.2159935}.

\leavevmode\vadjust pre{\hypertarget{ref-Howard2018-iv}{}}%
Howard, Matt C., and Michael E. Hoffman. 2018. {``{Variable-Centered},
{Person-Centered}, and {Person-Specific} Approaches: {W}here Theory
Meets the Method.''} \emph{Organizational Research Methods} 21 (4):
846--76. \url{https://doi.org/10.1177/1094428117744021}.

\leavevmode\vadjust pre{\hypertarget{ref-Joreskog:1970}{}}%
Jöreskog, Karl G. 1970. {``A General Method for Analysis of Covariance
Structures.''} \emph{Biometrika} 57 (2): 239--51.

\leavevmode\vadjust pre{\hypertarget{ref-McLachlan:Krishnan:2008}{}}%
McLachlan, G. J., and T. Krishnan. 2008. \emph{The {EM} Algorithm and
Extensions}. 2nd ed. Vol. 382. Wiley Series in Probability and
Statistics. Hoboken, NJ, USA: Wiley-Interscience.

\leavevmode\vadjust pre{\hypertarget{ref-McLachlan:Peel:2000}{}}%
McLachlan, G. J., and D. Peel. 2000. \emph{Finite Mixture Models}. Vol.
299. Wiley Series in Probability and Statistics. New York, NY, USA: John
Wiley \& Sons.

\leavevmode\vadjust pre{\hypertarget{ref-McLachlan2003}{}}%
McLachlan, G. J., D. Peel, and R. W. Bean. 2003. {``Modelling
High-Dimensional Data by Mixtures of Factor Analyzers.''}
\emph{Computational Statistics \& Data Analysis} 41: 379--88.

\leavevmode\vadjust pre{\hypertarget{ref-McNicholas2008}{}}%
McNicholas, P. D., and T. B. Murphy. 2008. {``{Parsimonious {G}aussian
mixture models}.''} \emph{Statistics and Computing} 18 (3): 285--96.

\leavevmode\vadjust pre{\hypertarget{ref-McParland2016}{}}%
McParland, D., and I. C. Gormley. 2016. {``{M}odel Based Clustering for
Mixed Data: {clustMD}.''} \emph{Advances in Data Analysis and
Classification} 10 (2): 155--69.

\leavevmode\vadjust pre{\hypertarget{ref-MoEClust2020}{}}%
Murphy, Keefe, and Thomas Brendan Murphy. 2020. {``Gaussian Parsimonious
Clustering Models with Covariates and a Noise Component.''}
\emph{Advances in Data Analysis and Classification} 14 (2): 293--325.
\url{https://doi.org/10.1007/s11634-019-00373-8}.

\leavevmode\vadjust pre{\hypertarget{ref-Murphy2021}{}}%
Murphy, Keefe, Thomas Brendan Murphy, Raffaella Piccarreta, and Isobel
Claire Gormley. 2021. {``Clustering Longitudinal Life-Course Sequences
Using Mixtures of Exponential-Distance Models.''} \emph{Journal of the
Royal Statistical Society: Series A (Statistics in Society)} 184 (4):
1414--51. \url{https://doi.org/10.1111/rssa.12712}.

\leavevmode\vadjust pre{\hypertarget{ref-Murphy2020}{}}%
Murphy, Keefe, Cinzia Viroli, and I. Claire Gormley. 2020. {``Infinite
Mixtures of Infinite Factor Analysers.''} \emph{Bayesian Analysis} 15
(3): 937--63.

\leavevmode\vadjust pre{\hypertarget{ref-Newton:Raftery:1994}{}}%
Newton, Michael A, and Adrian E Raftery. 1994. {``Approximate Bayesian
Inference with the Weighted Likelihood Bootstrap (with Discussion).''}
\emph{Journal of the Royal Statistical Society: Series B (Statistical
Methodology)} 56 (1): 3--48.

\leavevmode\vadjust pre{\hypertarget{ref-NylundGibsonChoi2018}{}}%
Nylund-Gibson, K., and A. Y. Choi. 2018. {``Ten Frequently Asked
Questions about Latent Class Analysis.''} \emph{Translational Issues in
Psychological Science} 4 (4): 440--61.

\leavevmode\vadjust pre{\hypertarget{ref-OHagan:Murphy:Scrucca:Gormley:2019}{}}%
O'Hagan, Adrian, Thomas Brendan Murphy, Luca Scrucca, and Isobel Claire
Gormley. 2019. {``Investigation of Parameter Uncertainty in Clustering
Using a {G}aussian Mixture Model via Jackknife, Bootstrap and Weighted
Likelihood Bootstrap.''} \emph{Computational Statistics} 34 (4):
1779--1813. \url{https://doi.org/10.1007/s00180-019-00897-9}.

\leavevmode\vadjust pre{\hypertarget{ref-mixture2022}{}}%
Pocuca, Nik, Ryan P. Browne, and Paul D. McNicholas. 2022.
\emph{{mixture}: {M}ixture Models for Clustering and Classification}.
\url{https://CRAN.R-project.org/package=mixture}.

\leavevmode\vadjust pre{\hypertarget{ref-Rstat}{}}%
R Core Team. 2023. \emph{R: {A} Language and Environment for Statistical
Computing}. Vienna, Austria: {R Foundation for Statistical Computing}.
\url{https://www.R-project.org/}.

\leavevmode\vadjust pre{\hypertarget{ref-Rosenberg:Beymer:Anderson:VanLissa:Schmidt:2018}{}}%
Rosenberg, Joshua M., Patrick N. Beymer, Daniel J. Anderson, Caspar J.
Van Lissa, and Jennifer A. Schmidt. 2018. {``Tidy{LPA}: {A}n {R} Package
to Easily Carry Out Latent Profile Analysis ({LPA}) Using Open-Source or
Commercial Software.''} \emph{Journal of Open Source Software} 3 (30):
978. \url{https://doi.org/10.21105/joss.00978}.

\leavevmode\vadjust pre{\hypertarget{ref-Rubin:1981}{}}%
Rubin, Donald B. 1981. {``The {B}ayesian Bootstrap.''} \emph{The Annals
of Statistics} 9 (1): 130--34.

\leavevmode\vadjust pre{\hypertarget{ref-Saqr2021}{}}%
Saqr, Mohammed, and Sonsoles López-Pernas. 2021. {``The Longitudinal
Trajectories of Online Engagement over a Full Program.''}
\emph{Computers \& Education} 175 (December): 104325.
\url{https://doi.org/10.1016/j.compedu.2021.104325}.

\leavevmode\vadjust pre{\hypertarget{ref-Saqr2022-fp}{}}%
---------. 2022. {``How {CSCL} Roles Emerge, Persist, Transition, and
Evolve over Time: {A} Four-Year Longitudinal Study.''} \emph{Computers
\& Education} 189: 104581.
\url{https://doi.org/10.1016/j.compedu.2022.104581}.

\leavevmode\vadjust pre{\hypertarget{ref-Saqr2023-if}{}}%
Saqr, Mohammed, Sonsoles López-Pernas, Satu Helske, and Stefan
Hrastinski. 2023. {``The Longitudinal Association Between Engagement and
Achievement Varies by Time, Students' Subgroups, and Achievement State:
{A} Full Program Study.''} \emph{Computers \& Education} 199: 104787.
\url{https://doi.org/10.1016/j.compedu.2023.104787}.

\leavevmode\vadjust pre{\hypertarget{ref-Scheidt2021-sg}{}}%
Scheidt, Matthew, Allison Godwin, Edward Berger, John Chen, Brian P
Self, James M Widmann, and Ann Q Gates. 2021. {``Engineering Students'
Noncognitive and Affective Factors: {G}roup Differences from Cluster
Analysis.''} \emph{Journal of Engineering Education} 110 (2): 343--70.
\url{https://doi.org/10.1002/jee.20386}.

\leavevmode\vadjust pre{\hypertarget{ref-Schwarz:1978}{}}%
Schwarz, G. 1978. {``Estimating the Dimension of a Model.''} \emph{The
Annals of Statistics} 6 (2): 461--64.
\url{https://doi.org/10.1214/aos/1176344136}.

\leavevmode\vadjust pre{\hypertarget{ref-Scrucca:etal:2016}{}}%
Scrucca, Luca, Michael Fop, T. Brendan Murphy, and Adrian E. Raftery.
2016. {``{mclust} 5: {C}lustering, Classification and Density Estimation
Using {G}aussian Finite Mixture Models.''} \emph{{The R Journal}} 8 (1):
205--33. \url{https://doi.org/10.32614/RJ-2016-021}.

\leavevmode\vadjust pre{\hypertarget{ref-mclust:book:2023}{}}%
Scrucca, Luca, Chris Fraley, T. Brendan Murphy, and Adrian E. Raftery.
2023. \emph{Model-Based Clustering, Classification, and Density
Estimation Using {mclust} in {R}}. The {R} Series. London, UK: Chapman
\& Hall/CRC Press. \url{https://doi.org/10.1201/9781003277965}.

\leavevmode\vadjust pre{\hypertarget{ref-Spearman:1904}{}}%
Spearman, C. 1904. {``{{`General Intelligence,'}} Objectively Determined
and Measured.''} \emph{The American Journal of Psychology} 15 (2):
201--92. \url{https://doi.org/10.2307/1412107}.

\leavevmode\vadjust pre{\hypertarget{ref-Tormanen2022-ux}{}}%
Törmänen, Järvenoja, Saqr, Malmberg, et al. 2022. {``A Person-Centered
Approach to Study Students' Socio-Emotional Interaction Profiles and
Regulation of Collaborative Learning.''} \emph{Frontiers in Education}
7. \url{https://doi.org/10.3389/feduc.2022.866612}.

\leavevmode\vadjust pre{\hypertarget{ref-Yu2022-fr}{}}%
Yu, Jianhui, Changqin Huang, Tao He, Xizhe Wang, and Linjie Zhang. 2022.
{``Investigating Students' Emotional Self-Efficacy Profiles and Their
Relations to Self-Regulation, Motivation, and Academic Performance in
Online Learning Contexts: {A} Person-Centered Approach.''}
\emph{Education and Information Technologies} 27 (8): 11715--40.
\url{https://doi.org/10.1007/s10639-022-11099-0}.

\leavevmode\vadjust pre{\hypertarget{ref-Zhang2023-zt}{}}%
Zhang, Yingbin, Luc Paquette, Juan D Pinto, Qianhui Liu, and Aysa Xuemo
Fan. 2023. {``Combining Latent Profile Analysis and Programming Traces
to Understand Novices' Differences in Debugging.''} \emph{Education and
Information Technologies} 28 (4): 4673--4701.
\url{https://doi.org/10.1007/s10639-022-11343-7}.

\leavevmode\vadjust pre{\hypertarget{ref-Zhen2019}{}}%
Zhen, Rui, Ru-De Liu, Ming-Te Wang, Yi Ding, Ronghuan Jiang, Xinchen Fu,
and Yan Sun. 2019. {``Trajectory Patterns of Academic Engagement Among
Elementary School Students: {T}he Implicit Theory of Intelligence and
Academic Self-Efficacy Matters.''} \emph{British Journal of Educational
Psychology} 90 (3): 618--34. \url{https://doi.org/10.1111/bjep.12320}.

\leavevmode\vadjust pre{\hypertarget{ref-Zucchini:MacDonald:Langrock:2016}{}}%
Zucchini, Walter, Iain L MacDonald, and Roland Langrock. 2016.
\emph{Hidden {M}arkov Models for Time Series: {A}n Introduction Using
{R}}. Vol. 105. Monographs on Statistics and Applied Probability.
London, UK: Chapman \& Hall/CRC Press.

\end{CSLReferences}

\end{document}